\documentclass[twocolumn]{aastex631}

\usepackage{newtxtext,newtxmath}

\usepackage[T1]{fontenc}
\usepackage{ae,aecompl}
\usepackage{graphicx}	% Including figure files
\usepackage{amsmath}	% Advanced maths commands
\usepackage{color}
\usepackage{here}
\usepackage{physics}

\newcommand{\hMpc}{$ \, h^{-1}  \rm Mpc$}

\newcommand{\Mpccube}{$\rm Mpc^3$}

\newcommand{\hpc}{$ \, h^{-1} \rm pc$}

\newcommand{\hMsun}{$\, h^{-1} \rm M_\odot$}
\newcommand{\Msun}{$\, \rm M_\odot$}
\newcommand{\mvir}{$M_{\rm vir}$}
\newcommand{\rvir}{$R_{\rm vir}$}
\newcommand{\mcrit}{$M_{\rm crit}$}

\newcommand{\mcold}{$M_{\rm cold}$}
\newcommand{\mhot}{$M_{\rm hot}$}
\newcommand{\mstar}{$M_*$}
\newcommand{\kms}{$\, \rm km \, s^{-1}$}
\newcommand{\lsim}{\mbox{${\,\hbox{\hbox{$ < $}\kern -0.8em \lower 1.0ex\hbox{$\sim$}}\,}$}}
\newcommand{\gsim}{\mbox{${\,\hbox{\hbox{$ > $}\kern -0.8em \lower 1.0ex\hbox{$\sim$}}\,}$}}
\newcommand{\p}{Pop III}
\newcommand{\pii}{Pop II}
\newcommand{\pandii}{Pop III and II}
\newcommand{\jlw}{$J_{\rm LW}$}
\newcommand{\vbc}{$v_{\rm bc}$}

\newcommand{\mpopiii}{$M_{\rm III}$}
\newcommand{\tvir}{$T_{\rm vir}$}
\newcommand{\rcut}{$r_{\rm cut}$}
\newcommand{\ndsms}{$n_{\rm SMS}$}
\newcommand{\nsms}{$\overline{N}_{\rm SMS}$}
\newcommand{\sv}{$\sigma_{\rm vbc}$}
\newcommand{\phif}{Phi-4096}
\newcommand{\Me}{M8}
\newcommand{\Mt}{M3}
\newcommand{\Ht}{H3}
\newcommand{\Lt}{L3}

\definecolor{mygray}{gray}{0.7}
\definecolor{darkgreen}{rgb}{0.0, 0.26, 0.15}
\begin{document}
\title[A semi-analytic framework of Pop III]
{A Semi-analytic Framework of Population III and Subsequent Galaxy Formation on Cosmological $N$-body Simulations}

\author[0000-0002-5316-9171]{Tomoaki Ishiyama}\affiliation{Digital Transformation Enhancement Council, Chiba University, 1-33, Yayoi-cho, Inage-ku, Chiba, 263-8522, Japan}
\author[0000-0002-4317-767X]{Shingo Hirano}\affiliation{Department of Applied Physics, Faculty of Engineering, Kanagawa University, 
3-27-1, Rokukakubashi, Kanagawa-ku, Yokohama-shi, Kanagawa 221-0802, Japan}
\correspondingauthor{Tomoaki Ishiyama}
\email{ishiyama@chiba-u.jp}

%%%%%%%%%%%%%%%%%%%%%%%%%%%%%%%%%%%%%%%%%%%%%%%%%%%%%%%%%%%%%%%%%%%%%%%%%%%%%%

\begin{abstract} 
We develop a new semi-analytic framework of Population (Pop) III and
subsequent galaxy formation designed to run on dark matter halo merger
trees.  In our framework, we consider the effect of the Lyman-Werner
flux from Pop III and II stars and the dark matter baryon streaming
velocity on the critical halo mass for the \p\ formation.  Our
model incorporates the Lyman-Werner feedback in a self-consistent way,
therefore, the spatial variation of Lyman-Werner feedback naturally emerges.
The \p\ mass depends on the properties of a halo as reproducing
radiative hydrodynamical simulation results. 
We perform statistical studies of \p\ stars by applying this framework
to high-resolution cosmological $N$-body simulations with a maximum
box size of 16\hMpc\ and enough mass resolution to resolve \p-forming halos.  
A top-heavy initial mass function emerges and two peaks
corresponding to the H$_2$ ($20 \lesssim z \lesssim 25$) and atomic
cooling halos ($z \lesssim 15$) exist in the distribution.
Supermassive stars can be formed in the atomic cooling halos, and the
fractions of such supermassive stars increase with the value of
streaming velocity.  At least an 8\hMpc\ simulation box and the
self-consistent model for the Lyman-Werner feedback are necessary to
correctly model the \p\ formation in the atomic cooling halos.  Our
model predicts one supermassive star per halo with several
$10^9$\Msun\ at z=$7.5$, which is enough to reproduce a high redshift
quasar.
\end{abstract}


%%%%%%%%%%%%%%%%%%%%%%%%%%%%%%%%%%%%%%%%%%%%%%%%%%%%%%%%%%%%%%%%%%%%%%%%%%%%%%
%%%%%%%%%%%%%%%%%%%%%%%%%%%%%%%%%%%%%%%%%%%%%%%%%%%%%%%%%%%%%%%%%%%%%%%%%%%%%%
%%%%%%%%%%%%%%%%%%%%%%%%%%%%%%%%%%%%%%%%%%%%%%%%%%%%%%%%%%%%%%%%%%%%%%%%%%%%%%
%%%%%%%%%%%%%%%%%%%%%%%%%%%%%%%%%%%%%%%%%%%%%%%%%%%%%%%%%%%%%%%%%%%%%%%%%%%%%%
\section{Introduction}\label{sec:intro}

Population III (\p) stars, also recognized as first stars, are the
first luminous objects in the Universe.  \p\ stars are born in
primordial gas within dark matter halos with masses ranging from
approximately $10^5$ to $10^7$ \Msun, around 100 million years after
the Big Bang \citep[e.g., ][]{Haiman1996, Nishi1999, Abel2002,
  Yoshida2003}.  The mass of \p\ stars is predicted to surpass that of
present-day stars \citep[e.g.,
][]{Omukai1998,Omukai2001,Omukai2003,Bromm2004} because the cooling
via H$_2$, which is the dominant coolant for the
primordial gas in such early Universe, is insufficient.  Cosmological radiation
hydrodynamical simulations have been consistently predicting that the
\p\ stars are typically very massive
\citep[e.g.,][]{Abel2002,Yoshida2006,Yoshida2008,Hosokawa2011,Susa2014,Hirano2014,Hirano2015,Toyouchi2023}.

Understanding the initial mass function (IMF) of \p\ stars is
essential because it has a large impact on subsequent various
phenomena, encompassing the formation of first galaxies and the seeds
of supermassive black holes, the cosmic reionization, and the
early metal enrichment of the Universe.  Consequently, the IMF can be
constrained by observations of high redshift galaxies and active
galactic nuclei \citep[AGN; e.g.,][for reviews and references
  therein]{Inayoshi2020}, and the spatial inhomogeneity of the
intergalactic medium \citep[IGM; e.g., ][]{Kirihara2020}.  In fact,
early observations conducted by the James Webb Space Telescope (JWST)
indicate a preference for top-heavy IMF \citep[e.g.,][]{Harikane2023a}, 
although it is still controversial.

Multiple different observations can be used to constrain specific models of the
  Pop III star formation process and the resulting IMF.
Fragmentation of circumstellar disks in the vicinity of \p\ stars can arise
due to the gravitational instability, giving rise to the birth of massive
\p\ binary stars \citep{Sugimura2020,Latif2022,Sugimura2023,Kirihara2023}.
These binary stars can evolve into progenitors of binary black holes
observable through gravitational waves \citep[e.g.,][]{Kinugawa2014}.
Low-mass \p\ stars can also form
\citep[e.g.,][]{Clark2008,Smith2011,Clark2011b,Clark2011,Greif2011,Greif2012,
  Machida2013,Susa2013,Susa2014,Susa2019,Nishijima2024}, and they
could survive beyond the current Hubble, if their mass falls below
0.8\Msun.  They can be detected in the Milky-Way as metal-free or
-poor stars.  Their abundance can constrain the low-mass \p\ IMF
\citep{Hartwig2015, Ishiyama2016} and the number density of
interstellar objects \citep{Kirihara2019}.

To facilitate detailed theoretical predictions for these observations,
we need to comprehensively understand statistics of \p\ stars at high
redshift, such as their abundance (for instance, the number
  density and the \p\ to II ratio), IMF, star formation rate, spatial
distribution, and their redshift evolution. Cosmological
hydrodynamical simulations and semi-analytic models on dark matter halo
merger trees have been utilized in this effort
\citep[e.g.,][]{Fialkov2012,Agarwal2012,Susa2014,Hirano2014,Hirano2015,
  Visbal2018, Visbal2020,Tanaka2021,Hartwig2022,Chiaki2023,Hedge2023,
  Bovill2024,Feathers2024,Trinca2024,Ventura2024}.  Nevertheless,
previous studies have employed relatively limited spatial volume, 
typically a few \hMpc\ (e.g., $\sim$3\hMpc\ for \citealt{Hirano2015,Visbal2018} and
  2\hMpc\ for \citealt{Visbal2020}), except for one with 10 \hMpc\ \citep{Ventura2024}
, thereby constraining the
statistical analyses due to the absence of large-scale
fluctuations. It is also quite difficult to capture the inhomogeneity
of Lyman-Werner (LW) background.

\begin{figure*}
\centering
\includegraphics[width=0.82\linewidth]{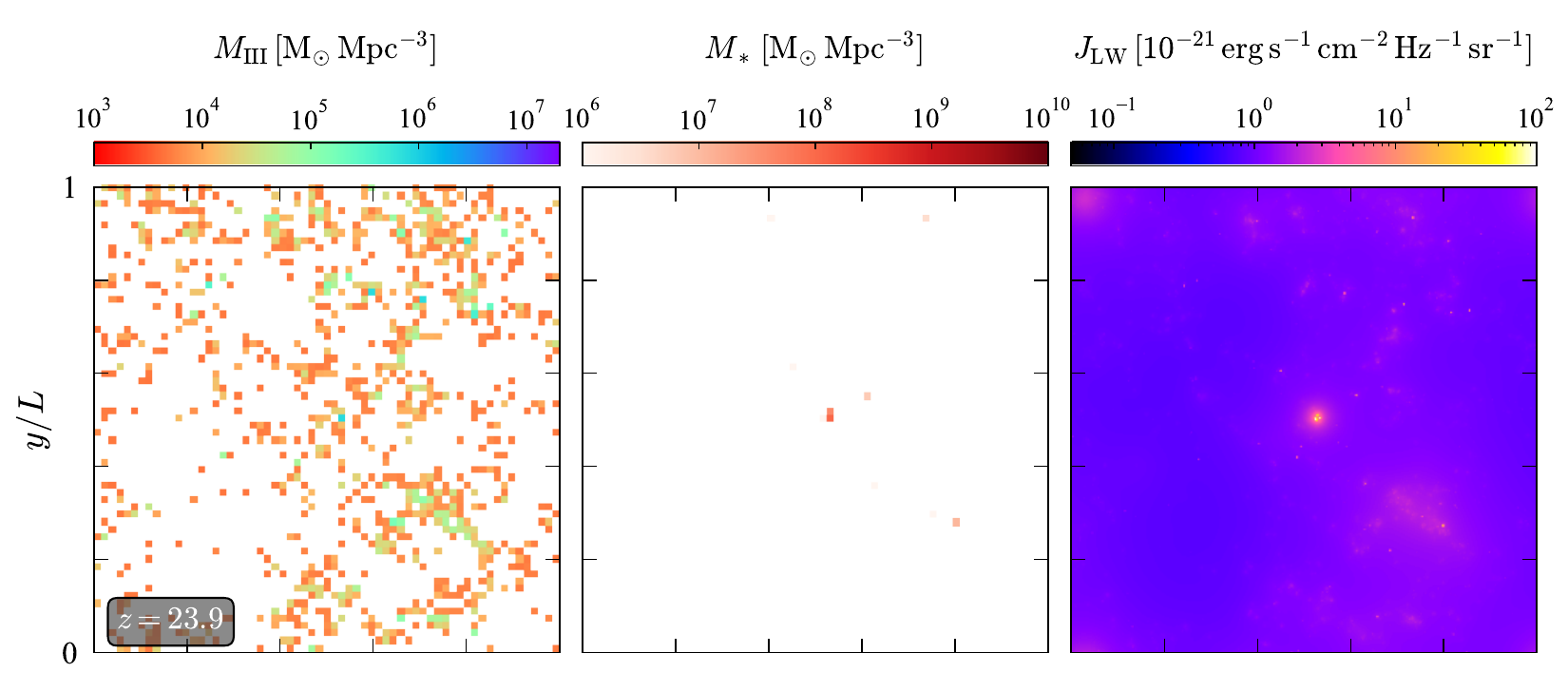}
\includegraphics[width=0.82\linewidth]{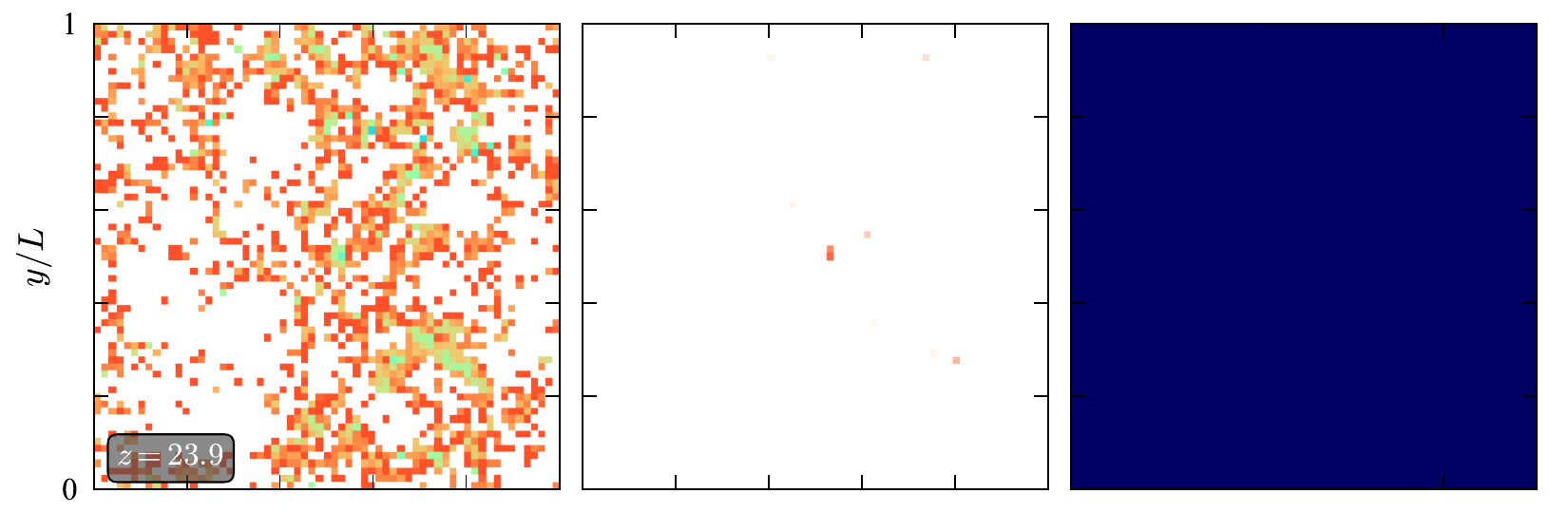}
\includegraphics[width=0.82\linewidth]{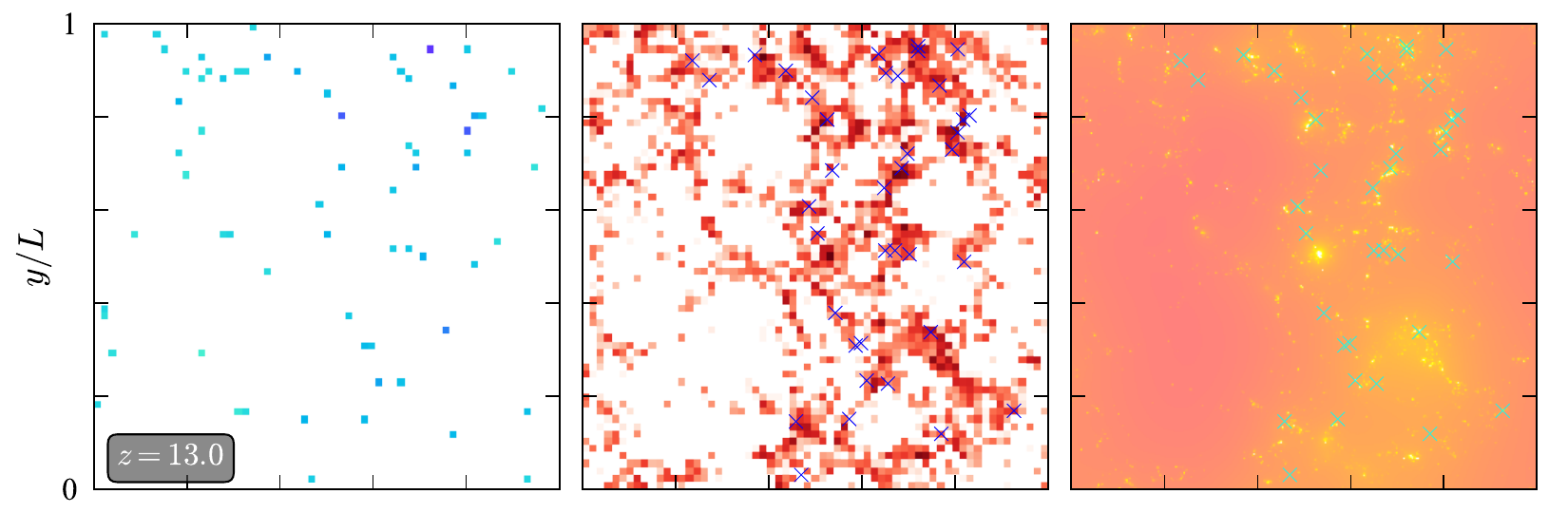}
\includegraphics[width=0.82\linewidth]{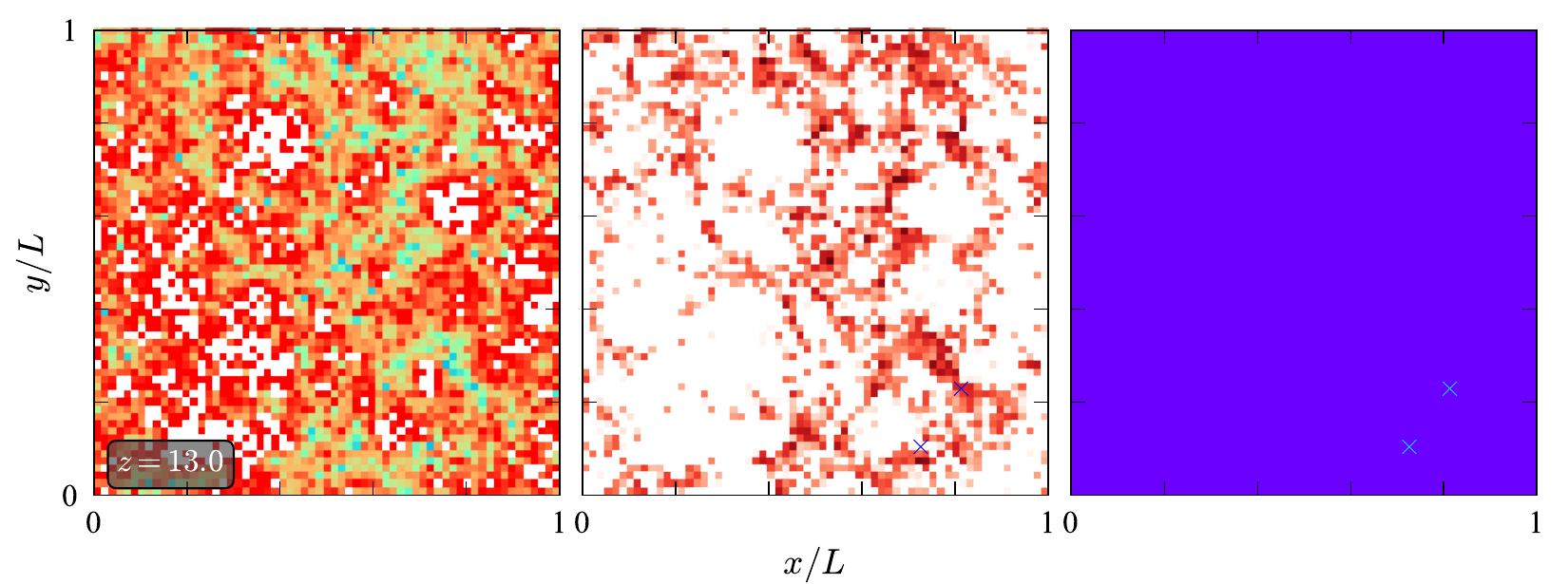}
\caption{
Difference of the self-consistent (first and third rows) and 
the uniform models (second and fourth rows) on the fiducial \phif\ simulation
($L=16$\hMpc) with zero streaming velocity. 
Supermassive \p\ stars (more massive than $10^{4.5}$\Msun) are displayed by crosses. 
From left to right, each panel shows the spatial distribution of the total 
mass density of \p, \pii\ stars, and Lyman-Werner flux, respectively.
See also Figure~\ref{fig:jlw}.
}
\label{fig:visual}
\end{figure*}

In \cite{Ishiyama2016}, we have constrained the low mass \p\ IMF
employing a simple semi-analytic model and a high-resolution
cosmological $N$-body simulation having enough resolution ($<
10^4$\Msun) tailored for the \p-forming halos and larger volume (8
\hMpc) compared to the others mentioned above.  In the present paper,
we use a cosmological $N$-body simulation having an even larger
volume (16 \hMpc) to perform statistical studies of \p\ stars. This
simulation was followed down to $z=0$, demonstrating the stark
contrast to simulations used in other semi-analytic models.  This
allows us to predict observational signals of \p\ stars in the current
Milky-Way focusing on the galactic archaeology. We also develop a
semi-analytic framework of \p\ and subsequent galaxy formation
designed to run on halo merger trees.  Motivated by results of
radiative hydrodynamical simulations, we model the critical halo
mass for the \p\ formation, its dependency on the LW flux from
\pandii\ sources throughout the entire volume and the dark matter
baryon streaming velocity, and \p\ IMF.  Our model incorporates LW
feedback in a self-consistent way, therefore, spatial information of
halos is mandatory.

The spatial inhomogeneities of LW intensity has a big impact on 
the statistical properties of \pandii\ stars. 
Figure~\ref{fig:visual} shows that the introductory visualization of 
model results. 
Compared to the uniform LW background model, 
the self-consistent model promotes the formation of supermassive \p\ stars (displayed by crosses)
and prevents that of less massive stars at later epoch.
The resulting initial mass function of \p\ stars significantly differs between both models
as shown in the remaining of the paper. 

In \S~\ref{sec:sim}, we describe the numerical detail of the
cosmological simulations employed in this study.  \S~\ref{sec:model}
gives the details of a new semi-analytic framework of \p\ formation
and subsequent galaxy formation.  We present results of model
calculations in \S~\ref{sec:result}, and scrutinize the impact of the
LW feedback model, as well as the box size of $N$-body simulations ,
on the statistical properties of \pandii\ stars at high redshift.
\S~\ref{sec:discussion} is dedicated to discussion, and
\S~\ref{sec:summary} is for the summary and future prospects.  In the
Appendix, we scrutinize the impact of the mass resolution on the
statistics.

%%%%%%%%%%%%%%%%%%%%%%%%%%%%%%%%%%%%%%%%%%%%%%%%%%%%%%%%%%%%%%%%%%%%%%%%%%%%%%
%%%%%%%%%%%%%%%%%%%%%%%%%%%%%%%%%%%%%%%%%%%%%%%%%%%%%%%%%%%%%%%%%%%%%%%%%%%%%%
%%%%%%%%%%%%%%%%%%%%%%%%%%%%%%%%%%%%%%%%%%%%%%%%%%%%%%%%%%%%%%%%%%%%%%%%%%%%%%
%%%%%%%%%%%%%%%%%%%%%%%%%%%%%%%%%%%%%%%%%%%%%%%%%%%%%%%%%%%%%%%%%%%%%%%%%%%%%%

\section{Cosmological $N$-body simulations}\label{sec:sim}

Hierarchical formation and assembly of dark matter halos are followed
with large cosmological $N$-body simulations having enough resolution
tailored for the formation of \p\ halos.  Subsequently, we model
the formation of \pandii\ stars on the merger trees of halos,
employing a semi-analytical approach.  The details of our
semi-analytic model are expounded upon in \S~\ref{sec:model}.

As a fiducial simulation, we adopt the \phif\ simulation
\citep{Ishiyama2021}, which was run with $4096^3$ dark matter
particles in a comoving box of 16\hMpc.  The particle mass resolution
and the gravitational softening length are $m_{\rm p} = 5.13 \times 10^{3}$\hMsun\ 
and $\varepsilon=60$\hpc, respectively. We additionally performed four smaller
simulations to investigate the effect of box size and resolution.  Two
of them, named \Me\ and \Mt, have the same mass and force resolution with
\phif, but the box sizes of \Me\ and \Mt\ are 8\hMpc\ and 3\hMpc,
respectively. The box size of 3\hMpc\ is also similar to those used in
previous works such as \citet{Hirano2015} ($\sim$ 3\hMpc) and
\citep{Visbal2020} ($\sim$ 2\hMpc), enabling easy comparison with each other.
The simulation \Ht\ is the highest resolution simulation
with $m_{\rm p} = 6.41 \times 10^{2}$\hMsun\ and $\varepsilon=30$\hpc.
The lowest resolution
simulation is named \Lt\ with $m_{\rm p} = 1.73 \times 10^{4}$\hMsun\ and $\varepsilon=90$\hpc.
This mass resolution is similar to those used in other 
semi-analytic studies \citep{Magg2018, Griffen2018}.
The initial seed of \Ht, \Mt, and \Lt\ is identical.
The basic parameters of those simulations are listed in Table~\ref{tab:sim}.

The initial conditions of the \phif\ were generated by the publicly
available code,
\textsc{music}~\footnote{\url{https://bitbucket.org/ohahn/music/}}
\citep{Hahn2011}.  
Those of all the other simulations were generated
by \textsc{2LPTic}
code~\footnote{\url{http://cosmo.nyu.edu/roman/2LPT/}}.  Both codes
adopt the second-order Lagrangian perturbation theory
\citep{Crocce2006}.  We calculated the matter transfer function using the
online version of
\textsc{Camb}~\footnote{\url{http://lambda.gsfc.nasa.gov/toolbox/tb\_camb\_form.cfm}}
\citep{Lewis2000}.  
The cosmological parameters of all simulations are
$\Omega_0=0.31$, $\Omega_{\rm b}=0.048$, $\lambda_0=0.69$, $h=0.68$,
$n_{\rm s}=0.96$, and $\sigma_8=0.83$, which are consistent with the
latest measurement by the Planck Satellite \citep{Planck2020}.
The initial redshift of all simulations is $z=127$.

\begin{table*}[t]
\centering
\caption
{Parameters of cosmological $N$-body simulations used in this study. 
Here, $N$, $L$, $\varepsilon$, $m_{\rm p}$, $z_{\rm fin}$, and $z_{\rm start,mrgt}$ are the total number of particles, box length, softening
  length, particle mass resolution, the final redshift of simulations, and the starting redshift of halo/subhalo merger trees, respectively. }
\label{tab:sim}
\begin{tabular}{lccccccc}
\hline
Name & $N$ & $L$(\hMpc) & $\varepsilon$ (\hpc) & $m_{\rm p}$ (\hMsun) & $z_{\rm fin}$ & $z_{\rm start,mrgt}$ \\
\hline 
\phif\ & $4096^3$ & 16.0 & 60 & $5.13 \times 10^{3}$ & 0.0 & 43\\
\Me\ & $2048^3$ & 8.0 & 60 & $5.13 \times 10^{3}$ & 7.5 & 31 \\
\Ht\ & $1536^3$ & 3.0 & 30  & $6.41 \times 10^{2}$ & 7.5 & 31 \\
\Mt\ & $768^3$  & 3.0 & 60  & $5.13 \times 10^{3}$ & 7.5 & 31 \\
\Lt\ & $512^3$  & 3.0 & 90  & $1.73 \times 10^{4}$ & 7.5 & 31 \\
\hline
\end{tabular}
\end{table*}

The gravitational evolution was followed by a massively parallel
TreePM code,
\textsc{GreeM}~\footnote{\url{http://hpc.imit.chiba-u.jp/~ishiymtm/greem/}}
\citep{Ishiyama2009b, Ishiyama2012} on Aterui II supercomputer at
Center for Computational Astrophysics, CfCA, of National Astronomical
Observatory of Japan.  We accelerated the gravity calculation using
\textsc{Phantom-grape}\footnote{\url{https://bitbucket.org/kohji/phantom-grape/src}}
code~\citep{Nitadori2006, Tanikawa2012, Tanikawa2013, Yoshikawa2018}.
Then, we constructed their merger trees using \textsc{Rockstar} phase space
halo/subhalo
finder~\footnote{\url{https://bitbucket.org/gfcstanford/rockstar/}}
\citep{Behroozi2013} and \textsc{consistent trees merger tree code}~\footnote{\url{https://bitbucket.org/pbehroozi/consistent-trees/}}
\citep{Behroozi2013b} for the entire volume of all simulations from
redshift $z_{\rm start,mrgt}$ to $\sim 7.5$
~\footnote{The halo/subhalo finding was stopped at z = 7.5, as
it became computationally prohibitive for some massive halos at low redshift.}
, where $z_{\rm
  start,mrgt}$ is the starting redshift of halo/subhalo merger trees.
For \phif, $z_{\rm start,mrgt} = 43$ 
when first halos emerge in the merger tree, 
and for the others, $z_{\rm
  start,mrgt} = 31$.  The number of snapshots stored is 58
between $z\sim31$ to $\sim7.5$ for all simulations, and the \phif\ additionally has
14 snapshots between $z\sim43$ to $\sim31$.  The
snapshot's logarithmic temporal interval $\varDelta \log(1+z)$ is about
0.01.

%%%%%%%%%%%%%%%%%%%%%%%%%%%%%%%%%%%%%%%%%%%%%%%%%%%%%%%%%%%%%%%%%%%%%%%%%%%%%%
%%%%%%%%%%%%%%%%%%%%%%%%%%%%%%%%%%%%%%%%%%%%%%%%%%%%%%%%%%%%%%%%%%%%%%%%%%%%%%
%%%%%%%%%%%%%%%%%%%%%%%%%%%%%%%%%%%%%%%%%%%%%%%%%%%%%%%%%%%%%%%%%%%%%%%%%%%%%%
%%%%%%%%%%%%%%%%%%%%%%%%%%%%%%%%%%%%%%%%%%%%%%%%%%%%%%%%%%%%%%%%%%%%%%%%%%%%%%
\section{Semi-analytic model}\label{sec:model}

In this section, we describe our semi-analytic model for the formation
of \pandii\ stars, and their Lyman-Werner (LW) radiation,
designed to run on the merger histories of dark matter halos taken
from cosmological $N$-body simulations.  In our model, the LW feedback
is modelled as a self-consistent way, hence the spatial information of
halos is mandatory.

%%%%%%%%%%%%%%%%%%%%%%%%%%%%%%%%%%%%%%%%%%%%%%%%%
%%%%%%%%%%%%%%%%%%%%%%%%%%%%%%%%%%%%%%%%%%%%%%%%%
\subsection{Sub-stepping}\label{sec:model:substep}

The temporal intervals between global timesteps of merger trees are
typically larger than the lifetime of massive \p\ stars.  For instance,
the intervals of global timesteps are about 6 Myr at $z=20$ and 16.6
Myr at $z=10$ in our merger trees, exceeding the lifetime of massive
stars \citep[1$\sim$10 Myr:][]{Schaerer2002}.  To track the star
formation precisely, we use sub-stepping technique of merger trees.  We
decompose each global timestep of merger trees into a constant number
of substeps, $N_{\rm sub}$, and randomly assign each halo merger
between global timesteps to one of these substeps.  We also compute
the dark matter smooth accretion mass ($M_{\rm smooth}$) between
global timesteps by subtracting the accumulated mass of all progenitor
halos from a given halo, and assume that the smooth mass accretion
occurs uniformly across substeps, as expressed by $dM = M_{\rm
  smooth}/N_{\rm sub}$.

By default, we set $N_{\rm sub} = 50$. In this case, sub-timesteps are
much less than and comparable to the lifetime of a massive star at high and
low redshift, respectively. This value also ensures the convergence of
\p\ statistics, such as the global star formation rate and the IMF.
As described in the following sections, the \p\ formation, the
baryon cycling, and the stellar LW feedback are calculated at each
substep.

%%%%%%%%%%%%%%%%%%%%%%%%%%%%%%%%%%%%%%%%%%%%%%%%%
%%%%%%%%%%%%%%%%%%%%%%%%%%%%%%%%%%%%%%%%%%%%%%%%%
\subsection{Condition of \p\ formation}\label{sec:model:pop3}
By tracking the merger trees of halos from the highest redshift, we
compare the virial mass of each halo \mvir\ with a critical value
\mcrit\ that depends on redshift, LW flux, and streaming velocity.  We
assume a single \p\ stars form in pristine cold dense gas in halos 
with $M_{\rm vir} \geq M_{\rm crit}$.  If any progenitor of a halo
experiences preceding \p\ formation, we assume that metal is provided
in the given halo and suppresses subsequent \p\ formation in the halo.
For \mcrit, we adopt a fitting function based on cosmological
hydrodynamical simulations proposed by \citet{Kulkarni2021}, which
takes the dependency of LW flux on each halo, the value of
streaming velocity, and redshift. 
In their simulations, the effect of H$_{2}$ self-shielding is taken into account.
This fitting function
gives the critical mass $M_{\rm K21}$ as
\begin{eqnarray}
  \label{eq:kulkarni21}
  M_{\rm K21}(J_{\rm LW},v_{\rm bc},z) &=& M_{z20} (J_{\rm LW}, v_{\rm bc}) \, {\rm M_\odot}\\ \nonumber
  && \times \, \qty( \frac{1+z}{21} )^{-\alpha_{\rm K21}(J_{\rm LW}, v_{\rm bc})},
\end{eqnarray}
where, $J_{\rm LW}$ and $v_{\rm bc}$ are the LW flux on a halo and the
streaming velocity in the units of $10^{-21} \, \rm erg \, s^{-1} \,
cm^{-2} \, Hz^{-1} \, sr^{-1}$ and \kms, respectively. $M_{z20}$ and
$\alpha_{\rm K21}$ are functional forms of $J_{\rm LW}$ and $v_{\rm
  bc}$ (see Eqs.~(3-12) of \citealt{Kulkarni2021} for the detail).  In our
semi-analytic model, we self-consistently calculate $J_{\rm LW}$ on
each halo following the procedure described in \S~\ref{sec:model:LW}.
The value of streaming velocity $v_{\rm bc}$ is a model parameter.  The
typical variance of the streaming velocity is $\sigma_{\rm vbc}=30$\kms\ 
at the time of recombination, and this is coherent across
several comoving Mpc \citep{Tseliakhovich2010, Tseliakhovich2011},
which is smaller than the box size of \phif\ simulation.
Nevertheless, we assume it is uniform across the entire volume to
easily compare how the different values of streaming velocity
impact on \p\ statistics.

\citet{Kulkarni2021} studied the critical halo mass using cosmological
hydrodynamical simulations with varying \jlw\ and \vbc. However, the
maximum values of \jlw\ studied by them are 30, 10 , and 1 for $v_{\rm
  bc} = 0, 1\sigma_{\rm vbc}$, and 2\sv, respectively.
We must pay attention when we apply this fitting function to outside of
these parameter spaces. In fact, when \vbc\ is high, 
this fitting function provides a smaller critical mass 
in higher \jlw\ ($>100$) than in smaller \jlw.
This is physically strange but is not surprising because
this fitting function was not calibrated for such high \jlw. To apply
this fitting function to such parameter spaces, we modify it
as follows, 
\begin{eqnarray}
  \label{eq:kulkarni21b}
  \max_{x \in [0,J_{\rm LW}]}{{M_{\rm K21}(x,v_{\rm bc},z)}} \to M_{\rm K21}(J_{\rm LW},v_{\rm bc},z). 
\end{eqnarray}
This recipe ensures that the critical mass increases monotonically
with \jlw.  In the Appendix~\ref{sec:mcrit_comp}, we show the
  comparison of \mcrit\ between the original model by
  \citet{Kulkarni2021} and our modification.

As \p\ and subsequent \pii\ star formation proceed, UV radiation in
the Lyman-Werner bands from these stars photodissociates H$_2$
molecules in relatively low mass halos  and suppresses \p\ formation
via H$_2$ cooling within the halos. If such pristine halos 
are massive enough (virial temperature \tvir\ $\gtrsim 8000$ K), atomic
hydrogen could be a major coolant and massive \p\ can form
\citep[e.g.,][]{Omukai2001,Bromm2003,Inayoshi2014,Ferandez2014,Becerra2015,Chon2016}.
We also follow such massive \p\ formation via atomic
cooling by adopting the following criterion of atomic cooling halo,
which was employed in \citet{Visbal2020} and \citet{Feathers2024} based on
cosmological hydrodynamical simulations of \citet{Ferandez2014},
\begin{eqnarray}
  M_{\rm a} = 5.4 \times 10^7 \, {\rm M_\odot} \qty(\frac{1+z}{11})^{-\frac{3}{2}}. 
  \label{eq:atomic_cooling}
\end{eqnarray}
Finally, we use the following expression as a critical halo mass of the \p\ formation, 
\begin{eqnarray}
  M_{\rm crit} = \min{(M_{\rm K21}, M_{\rm a})}.
  \label{eq:mcrit}
\end{eqnarray}

\begin{figure}
\centering
\includegraphics[width=\linewidth]{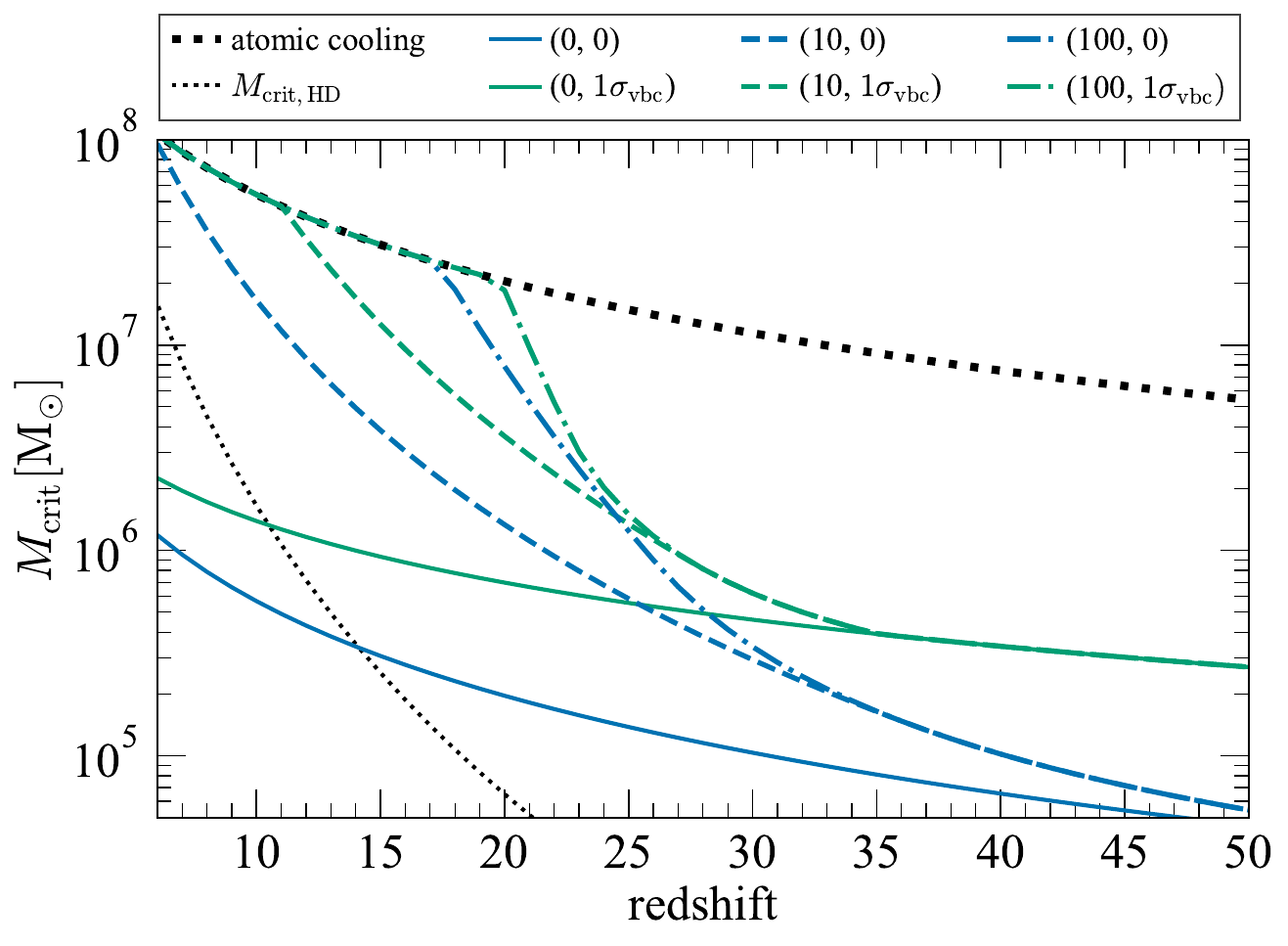}
\caption{
  Critical halo mass \mcrit\ of \p-forming halos as a function of redshift employed in our semi-analytic model [Eq~\eqref{eq:mcrit}].
  Solid, dashed, and dot-dashed curves indicate examples for various combinations of $(J_{\rm LW}, v_{\rm bc})$. 
Upward thick- and downward thin-dotted curves represent the critical mass of atomic cooling halos [Eq~\eqref{eq:atomic_cooling}] 
and the critical mass of the HD cooling [Eq~\eqref{eq:hd_cooling}], respectively.
}
\label{fig:mcrit}
\end{figure}

Figure~\ref{fig:mcrit} shows examples of critical halo mass \mcrit\ of
\p-forming halos as a function of redshift for $(J_{\rm LW}, v_{\rm
  bc}) = (0,0), (10,0), (100,0), (0,1\sigma_{\rm
  vbc}),(10,1\sigma_{\rm vbc})$, and~$(100,1\sigma_{\rm vbc})$.  The
critical mass gradually increases with decreasing redshift in any
case.  The existence of LW flux and streaming velocity further
increases the critical mass as expected from their physical
nature. The critical mass follows the atomic cooling limit
[Eq~\eqref{eq:atomic_cooling}] after redshift when the critical mass
described in Eqs.~\eqref{eq:kulkarni21}~and~\eqref{eq:kulkarni21b}
reaches this limit. This redshift is earlier in the higher streaming
velocity model than in the lower one because of the systematically
higher critical mass.

%%%%%%%%%%%%%%%%%%%%%%%%%%%%%%%%%%%%%%%%%%%%%%%%%
%%%%%%%%%%%%%%%%%%%%%%%%%%%%%%%%%%%%%%%%%%%%%%%%%
\subsection{\p\ mass}\label{sec:model:pop3_imf}
Once the virial mass of a pristine halo exceeds the critical mass
\mcrit\ described in the previous section, we assume that a single
\p\ star forms instantaneously in the halo. 
Note that recent radiative hydrodynamical simulations showed
that massive \p\ binary stars can be born as a result of fragmentation of
circumstellar disks in the vicinity of \p\ due to the gravitational
instability \citep{Sugimura2020,Sugimura2023}.
We discuss the effect of \p\ multiplicity on our model results 
in \S~\ref{sec:discussion:multiple}.

Previous studies assume that
the total mass of \p, $M_{\rm III}$, is constant in each halo
\citep{Feathers2024}, or is in proportion to the halo mass 
with a constant star formation efficiency \citep{Visbal2018,Visbal2020}.  In the case of
\citet{Hartwig2022}, a power-law IMF, 
$\frac{dN}{d\log{M_{\rm III}}} \propto M_{\rm III}^{-\alpha_{\rm III}}$,
 is used to assign the masses of \p\ stars
in a given halo, where $\alpha_{\rm III}$ is a free parameter.

Modern studies based on high-quality radiative hydrodynamical
simulations have shown that the mass of \p\ star is affected by
physical properties of host halos.  Using two-dimensional axisymmetric
radiative hydrodynamical simulations of \p\ formation in one hundred
halos sampled from cosmological simulations, \citet{Hirano2014} found
that the final \p\ stellar mass correlates with the physical
properties of the star-forming cloud and host halo, such as the gas
infall rate at the Jeans scale, $\dot{M}_{\rm Jeans}$.
\citet{Hirano2015} further extended this study and constructed an
empirical formula that connects $\dot{M}_{\rm Jeans}$ and the
evolution of halo virial mass as follows.
\begin{eqnarray}
    \label{eq:mdotjeans}
    \dot{M}_{\rm Jeans} = 
    \begin{cases}
      3 \times \dot{M}_{\rm vir} &(M_{\rm vir} > M_{\rm crit,HD}), \\
      0.3 \times \dot{M}_{\rm vir} &(M_{\rm vir}< M_{\rm crit,HD}), 
    \end{cases}
    \end{eqnarray}
where $\dot{M}_{\rm vir}$ is the mass accretion rate at virial scale empirically fitted by
\begin{eqnarray}
  \label{eq:macc}
  \dot{M}_{\rm vir} &=& 1.1 \times 10^{-3} \, {\rm M_\odot yr^{-1}} \qty(\frac{1+z}{20})^{3.5} \\ \nonumber && \times \, \qty(\frac{M_{\rm vir}}{4\times 10^5 \, {\rm M_\odot}})^{1.75}, 
\end{eqnarray}
and HD cooling is dominant rather than H$_2$ below $M_{\rm crit,HD}$, which is a critical value described as
\begin{eqnarray}
  \label{eq:hd_cooling}
  M_{\rm crit,HD} = 3.5 \times 10^5 \, {\rm M_\odot} \, \qty(\frac{1+z}{15})^{-5}.
\end{eqnarray}
The effect of HD cooling is not important for the critical halo mass given by Eq.~\eqref{eq:mcrit} because it does not operate from the very beginning of gas collapse. The collapse initially proceeds under H$_2$/atomic-cooling alone (at $n \sim 10^2\,{\rm cm^{-3}}$, where $n$ is the hydrogen number density). As the gas density increases ($n \sim 10^{5-6}\,{\rm cm^{-3}}$), HD cooling becomes efficient, leading to a further temperature decrease and a reduction in the accretion rate.
Therefore, the initial collapse condition of the gas within a halo is determined solely by H$_2$/atomic cooling, and Eq.~\eqref{eq:mcrit} is sufficient for this purpose. On the other hand, whether HD cooling becomes effective in a halo/gas cloud that begins collapse at the critical mass
requires an additional criterion that are described by Eqs.~(\ref{eq:mdotjeans}-\ref{eq:hd_cooling}).
Finally, the \p\ mass $M_{\rm H15,III}$ is modelled as a function of $\dot{M}_{\rm Jeans}$, 
\begin{eqnarray}
  M_{\rm H15,III} = 250 \, {\rm M_\odot} \qty(\frac{\dot{M}_{\rm Jeans}}{2.8 \times 10^{-3} \, {\rm M_\odot yr^{-1}}})^{0.7}.
\end{eqnarray}
We adopt this formula as a first approximation of the \p\ mass.

We also implement the supermassive star formation in our semi-analytic
model. \citet{Toyouchi2023} performed three-dimensional radiation
hydrodynamical simulations and found that supermassive stars can form
in atomic cooling halos. When the mass accretion rate is high enough,
$\dot{M}_{\rm vir} > 10^{-2} \, {\rm M_\odot yr^{-1}}$, nearly ten
times more massive \p\ stars could form (estimated from Figure~10 of
\citealt{Toyouchi2023}). Note that they obtained consistent
results with \citet{Hirano2014,Hirano2015} in the case of lower
accretion rate, $\dot{M}_{\rm vir} < 10^{-2} \, {\rm M_\odot
  yr^{-1}}$.  Hence, we populate a single \p\ star per halo
following the equation,
\begin{eqnarray}
  M_{\rm III} = 
\begin{cases}
  M_{\rm H15,III} &(\dot{M}_{\rm Jeans} \lesssim 10^{-2} \, {\rm M_\odot yr^{-1}}),\\
  10 \times M_{\rm H15,III} &(10^{-1} \, {\rm M_\odot yr^{-1}} \lesssim \dot{M}_{\rm Jeans}). 
\end{cases}
\end{eqnarray}
For $10^{-2} < \dot{M}_{\rm Jeans} < 10^{-1} \, {\rm M_\odot yr^{-1}}$, 
we assign $M_{\rm III}$ by log-linearly interpolating between 
$\eval{M_{\rm III}}{\dot{M}_{\rm Jeans}=10^{-2}}$ and $\eval{M_{\rm III}}{\dot{M}_{\rm Jeans}=10^{-1}}$. 

Once we assign the mass of a \p\ star, we determine the lifetime of the star
assuming no mass loss
using a table given by \citet{Schaerer2002}, which is based on realistic
population synthetic models. We linearly interpolate the table and
assume the lifetime of a \p\ star with mass more massive than $10^3 \, \rm
M_\odot$ being the same with that of $10^3 \, \rm M_\odot$ stars,
which is the upper limit of the table.  The UV radiation of
each \p\ star works with constant intensity within its lifetime 
(details are given in \S~\ref{sec:model:LW}).

%%%%%%%%%%%%%%%%%%%%%%%%%%%%%%%%%%%%%%%%%%%%%%%%%
%%%%%%%%%%%%%%%%%%%%%%%%%%%%%%%%%%%%%%%%%%%%%%%%%
\subsection{Baryon cycle}\label{sec:model:baryon}

Once the supernova explosion of a massive \p\ star occurs, metal
provided by the star mixes with the remaining pristine gas in the
halo.  
As shown in \S~\ref{sec:result}, the \p\ mass in a halo is always
predicted to exceed the upper limit mass of the pair-instability
supernova ($\sim 260$\Msun) for $v_{\rm bc} \gtrsim 1\sigma_{\rm
  vbc}$, therefore, metal enrichment by the supernova explosion does
not occur if we assume the spatially uniform streaming velocity across
the entire box,
no mass loss, and the absence of less massive \p\ stars than the
model prediction.
However, less massive \p\ stars must form around such massive \p\ stars in principle
because the coherent scale of the streaming velocity is several comoving Mpc. 
Therefore, we assume that these halos are massive enough to accrete nearby less massive halos  
that are already enriched, 
which do not form in the models with non-zero streaming velocity but must form in the real Universe, and the metal enrichment occurs in these halos to take the LW feedback from \pii\ into account.

If any progenitor of a halo is a halo that experiences preceding
\p\ formation, we assume that the subsequent \p\ formation in the given halo
is suppressed and the \pii\ formation occurs when it grows massive enough
for efficient gas cooling.  We take a critical value of a minimum
virial mass for such \pii\ formation halos to be $M_{\rm th}=10^7$\Msun\ as a fiducial value.  
We model the baryon cycle and the \pii\ formation in such halos by largely following a recipe used in 
\citet{Agarwal2012} and \citet{Chon2016} as follows. 

We consider three components of baryonic matter: non star-forming hot
gas \mhot, star-forming cold gas \mcold, and the total mass of
\pii\ stars \mstar.  We first place $M_{\rm hot} = f_{\rm b}M_{\rm vir}$ on each halo, 
where $f_{\rm b} = \Omega_{\rm b}/\Omega_{\rm 0}$, 
when they first emerge in the merger trees.  Then, we update each
component between substeps as follows.
%%%%%%%%%%%%%%%%%%%%%%%%%%%%%%%%%%%%%%%%%%%%%%%%%
\begin{enumerate}
\item {\it Smooth accretion of gas}. When smooth mass accretion occurs on a
  halo, the halo also accretes smooth gas. We assume that the gas is
  heated to the virial temperature, then add $f_{\rm b} \, dM$ to the
  hot gas component.
\item {\it Radiative gas cooling}. The radiative cooling of hot gas occurs 
  over the timescale of the halo dynamical time $t_{\rm dyn}$, which
  is given by
\begin{eqnarray}
  t_{\rm dyn} &=& \frac{R_{\rm vir}}{V_{\rm c}}, \\
  V_{\rm c} &=& 23.4 \qty[ \frac{\Omega_{\rm 0}}{\Omega_{\rm m}(z)} \frac{\Delta_{\rm c}}{18\pi^2}]^{\frac{1}{6}} \qty(\frac{M_{\rm vir}}{10^8 h^{-1} \rm M_\odot})^{\frac{1}{3}} \\ \nonumber
  && \times \, \qty(\frac{1+z}{10})^{\frac{1}{2}} \rm km \, s^{-1},
\end{eqnarray}
where, \rvir\ is the halo virial radius, $V_{\rm c}$ is the circular
velocity, $\Omega_{\rm m}(z)$ is the density parameter, and
$\Delta_{\rm c}=18\pi^2$ is the critical halo overdensity.
\item {\it Star formation}. In the metal enriched star-forming cold gas, new
  \pii\ stars are born over the timescale of
\begin{eqnarray}
  t_{\rm SF} &=& \frac{0.1 t_{\rm dyn}}{\alpha}
\end{eqnarray}
\citep[e.g.,][]{Kauffmann1999}, where $\alpha$ is the star formation efficiency and we set $\alpha=0.003$. 
\item {\it Supernovae feedback}. We assume a part of cold gas is converted
  into hot gas by supernovae explosions and the rate is in proportion
  to the star formation rate as 
\begin{eqnarray}
\gamma \frac{dM_{\rm *}}{dt} \equiv \gamma \frac{M_{\rm cold}}{t_{\rm SF}},
\end{eqnarray}
where
\begin{eqnarray}
  \gamma &=& \qty (\frac{V_{\rm c}}{V_{\rm out}})^{-\beta}
\end{eqnarray}
\citep{Cole1994}. We set $\beta=1.74$ and $V_{\rm out}=110 \,$\kms.
This process can be modelled as an outflow where the heated cold gas is
removed from the halo \citep{Agarwal2012}.  In our model, we do not
consider outflows and simply add the heated cold gas into the hot gas
component \citep{Chon2016}.
We consider this supernovae feedback from \pii\ stars, but not \p\ stars.
\end{enumerate}
%%%%%%%%%%%%%%%%%%%%%%%%%%%%%%%%%%%%%%%%%%%%%%%%%

The four processes are summarized in the following differential equations. 
\begin{eqnarray}
  \frac{dM_{\rm hot}}{dt} &=& -\frac{M_{\rm hot}}{t_{\rm dyn}} + \gamma \frac{M_{\rm cold}}{t_{\rm SF}} + f_{\rm b}\frac{dM_{\rm vir}}{dt}, \\ \nonumber
  \frac{dM_{\rm cold}}{dt} &=&\frac{M_{\rm hot}}{t_{\rm dyn}} - (1+\gamma) \frac{M_{\rm cold}}{t_{\rm SF}}, \\ \nonumber
  \frac{dM_{\rm *}}{dt} &=& \frac{M_{\rm cold}}{t_{\rm SF}}.
\end{eqnarray}
We explicitly integrate these equations on each substep. When a halo
merger occurs at a certain substep, these three components of the less
massive halo are added to those of the more massive halo.  We do not
follow the evolution of less massive halo at later steps.

%%%%%%%%%%%%%%%%%%%%%%%%%%%%%%%%%%%%%%%%%%%%%%%%%
%%%%%%%%%%%%%%%%%%%%%%%%%%%%%%%%%%%%%%%%%%%%%%%%%
\subsection{Lyman-Werner feedback}\label{sec:model:LW}

As described in \S~\ref{sec:model:pop3}, once \pandii\ stars
form, their UV radiation in the LW bands photodissociates H$_2$
molecules in relatively less massive and nearby halos  and
suppresses subsequent \p\ formation via H$_2$ cooling within the
halos.  If these halos grow massive enough, they could be a
possible site of massive \p\ formation via atomic cooling.  How the LW
radiation affects the critical mass of \p-forming halos  is
given in Eqs.~\eqref{eq:kulkarni21}~and~\eqref{eq:kulkarni21b}.  In
this section, we describe how to calculate the LW flux, \jlw, on each
halo.

\begin{figure*}
\centering
\includegraphics[width=0.55\linewidth]{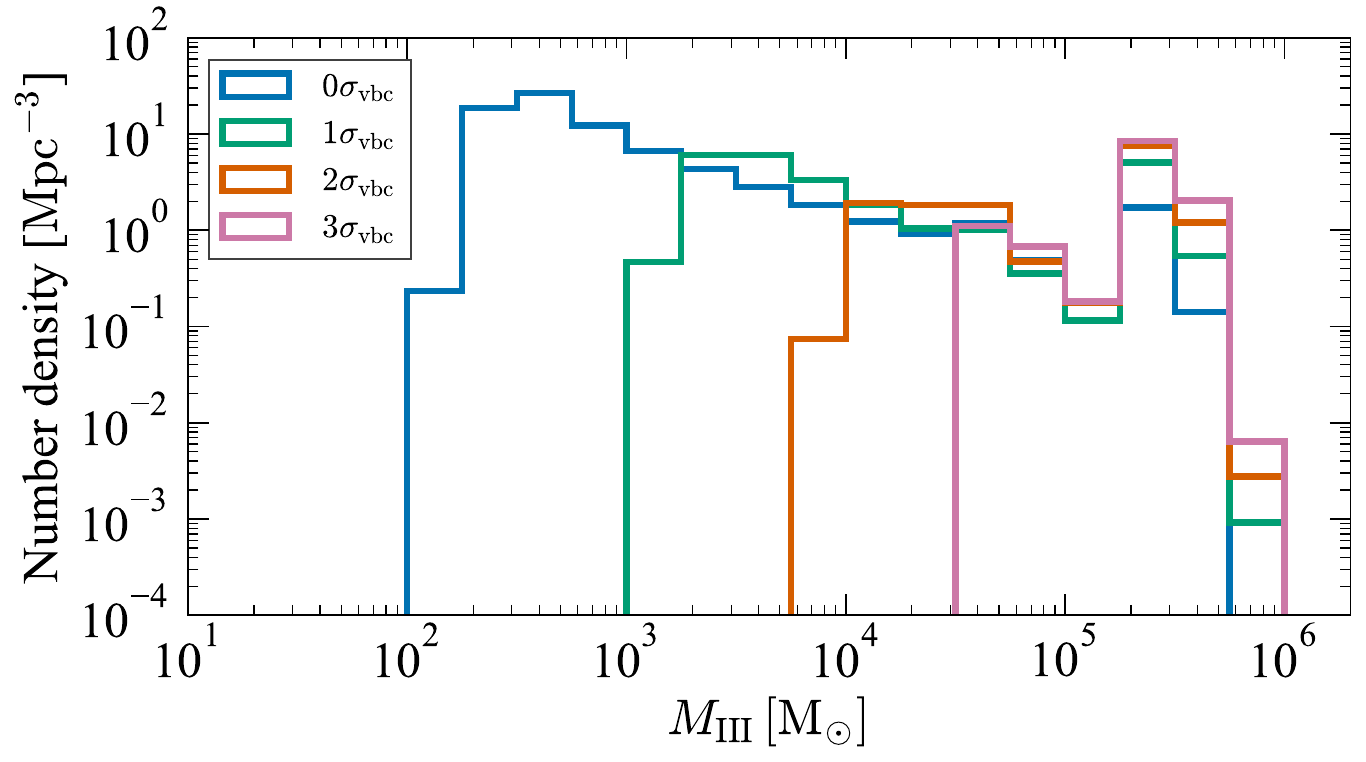}
\includegraphics[width=0.39\linewidth]{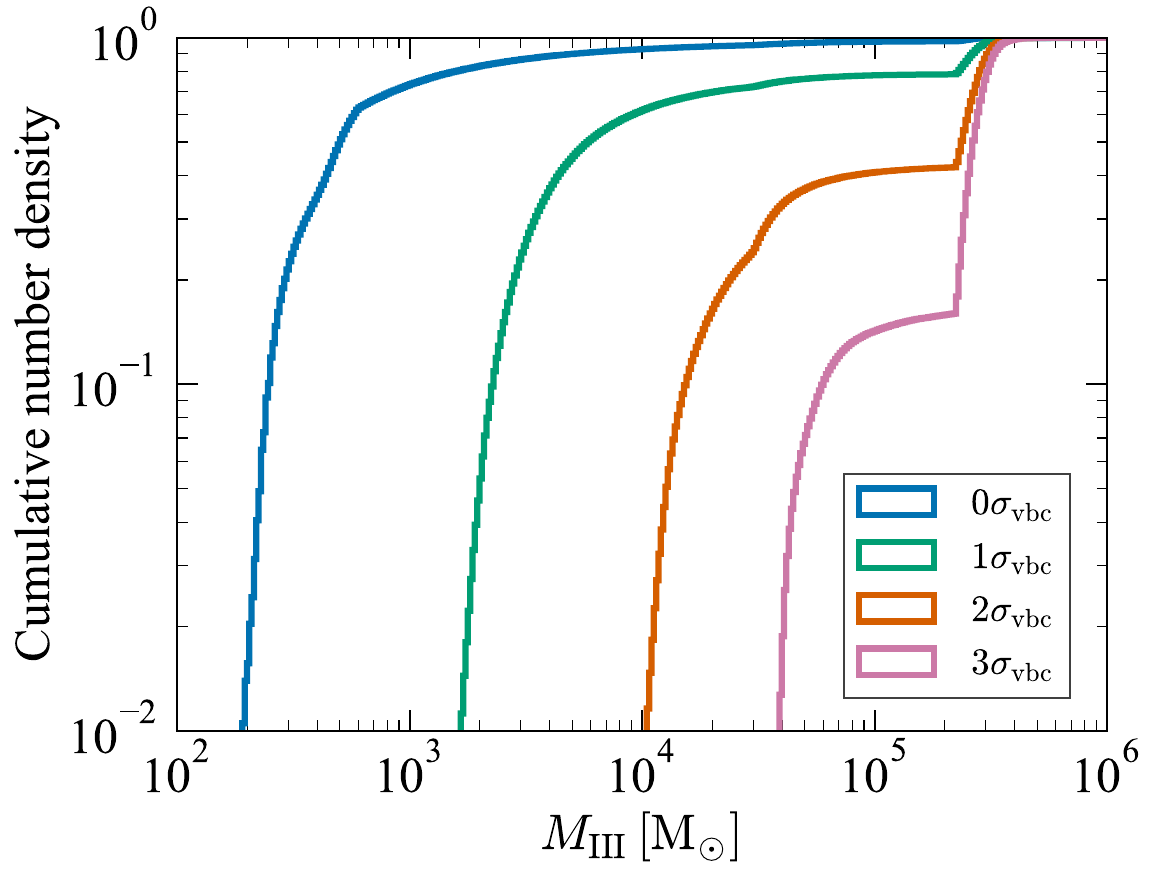}
\caption{ (Left) Number density of \p\ stars formed at all redshift as a
  function of the \p\ mass in a halo. We divide a single-digit
  mass range into four logarithmic bins and calculate the number
  density in each bin.  Four model results with the different values of streaming
  velocity, $v_{\rm bc}=0$, 1\sv, 2\sv, and
  3\sv, are shown.  (Right) Cumulative probability distribution of the 
number of \p\ stars as a function of the \p\ mass in a halo.
}
\label{fig:imf}
\end{figure*}

We adopt a self-consistent model for the LW feedback in which the LW
flux on each halo is calculated by accumulating the contributions from
all radiative sources of both living \pandii\ stars, resulting in
a spatially and time variable treatment. We employ the following
equations proposed by \citet{Johnson2013} to compute the
contribution from a source on a halo,
\begin{eqnarray}
  \label{eq:Johnson2013}
  J_{\rm LW,III} = 15 \qty(\frac{r}{\rm 1 \, kpc})^{-2} \qty( \frac{M_{\rm III}}{10^3 \, \rm M_\odot}), \\ \nonumber
  J_{\rm LW,II} = 3 \qty(\frac{r}{\rm 1 \, kpc})^{-2} \qty( \frac{M_{\rm *}}{10^3 \, \rm M_\odot}), 
\end{eqnarray}
where $r$ is the distance between 
the center of the source and that of a given halo in physical coordinates, 
where the center is defined as the most bound place in the halo.
The first equation gives the LW flux from a
\p\ star, while the second one is for a \pii-forming halo. For
$M_{\rm *}$, we instead use the total mass of \pii\ stars formed
within 5 Myr from the current substep, which is derived by multiplying
the instantaneous star formation rate at the substep with this period.

As the \pandii\ formation proceeds, the number of sources
increases. Therefore, the direct summation of
Eq~\eqref{eq:Johnson2013} is impractical because the calculation cost
is in proportion to both the number of sources and halos in the merger
tree. We accelerate this calculation using Fast Fourier Transformation
(FFT) as an approach in \citet{Visbal2020}. The periodic boundary
condition used in the cosmological simulations is also naturally
solved by the FFT. First, we set a uniform grid on the entire volume,
where the number of grid points is expressed by $N_{\rm g}$ in one
dimension.  We calculate the source energy (corresponds to a term
without $r^{-2}$ in Eq~\eqref{eq:Johnson2013}) from each grid by
assigning the energy of all sources nearby the given grid using the
CIC (Cloud in Cell) scheme.  Then, the LW flux from a grid to all other
grids is obtained using the FFT and the convolution with the Fourier
transformation of the $r^{-2}$ kernel. Finally, we calculate the LW
flux on each halo by interpolating the flux on the nearby grids.

The method described above is in analogy with the Particle-Mesh (PM)
method generally used in gravitational $N$-body simulations.  To
calculate the contribution of nearby sources more precisely, we adopt
a method in analogy with the Particle-Particle Particle-Mesh ($\rm
P^3M$) method, where we introduce the cutoff radius \rcut. The
contributions from all sources within the cutoff radius to a halo are
computed directly using Eq~\eqref{eq:Johnson2013}. For the
convolution, we set the kernel value to zero from the contributions
within the cutoff radius to avoid double counting. We set $N_{\rm
  g}=512^3, 256^3$, and $128^3$ for the simulations with the box size of
$L=16.0$, 8.0, and 3.0 \hMpc, respectively, and $r_{\rm cut} = L/N_{\rm
  g}$.  These values ensure the convergence of \p\ statistics. 
In terms of the number of \p\ stars formed at all redshift,
the difference between $N_{\rm g}=512^3$ and $256^3$ for the 16.0 \hMpc\ box 
is within 2\% level.

%%%%%%%%%%%%%%%%%%%%%%%%%%%%%%%%%%%%%%%%%%%%%%%%%%%%%%%%%%%%%%%%%%%%%%%%%%%%%%
%%%%%%%%%%%%%%%%%%%%%%%%%%%%%%%%%%%%%%%%%%%%%%%%%%%%%%%%%%%%%%%%%%%%%%%%%%%%%%
%%%%%%%%%%%%%%%%%%%%%%%%%%%%%%%%%%%%%%%%%%%%%%%%%%%%%%%%%%%%%%%%%%%%%%%%%%%%%%
%%%%%%%%%%%%%%%%%%%%%%%%%%%%%%%%%%%%%%%%%%%%%%%%%%%%%%%%%%%%%%%%%%%%%%%%%%%%%%
\section{Results}\label{sec:result}

We perform four semi-analytic calculations on the fiducial
\phif\ simulation with different values of the streaming velocity,
$v_{\rm bc}=0, 1\sigma_{\rm vbc}, 2\sigma_{\rm vbc}$, and
$3\sigma_{\rm vbc}$, where $\sigma_{\rm vbc} = 30$\kms. Other model
parameters are fixed across four calculations and shown in the
previous section. We start the calculations from $z=43$ when first
halos emerge in the merger tree and follow the evolution until
$z=7.5$.  In this section, we begin by showing the initial mass
function of \p\ formed at all redshift and compare the results between four
runs. Then, we compare the star formation history of these runs. We
also investigate the effect of box size (\S~\ref{sec:box_convergence}) 
and mass resolution (Appendix~\ref{sec:mass_convergence}) on 
statistical results using slim simulations.

%%%%%%%%%%%%%%%%%%%%%%%%%%%%%%%%%%%%%%%%%%%%%%%%%%%%%%%%%%%%%%%%%%%%%%%%%%%%%%
%%%%%%%%%%%%%%%%%%%%%%%%%%%%%%%%%%%%%%%%%%%%%%%%%%%%%%%%%%%%%%%%%%%%%%%%%%%%%%

\subsection{Initial mass function}\label{sec:imf}

The left panel of Figure~\ref{fig:imf} shows the number density of
\p\ stars formed at all redshift as a function of the \p\ mass within each
halo.  In all models, there are no halos  that contain \p\ with
\mpopiii\ $\lesssim 100$\Msun.  The critical halo mass of \p-forming
halos increases with decreasing redshift, while the mass accretion
rate increases with redshift and halo mass as shown in
  Eq~\eqref{eq:macc}.  This balance prevents the formation of less
massive \p\ stars.  The least mass of \p\ increases with the streaming
velocity as a natural consequence.  In all models, the number density
decreases with increasing \p\ mass from the first peak around the less
massive end and has the second peak at $M_{\rm III}=10^{(5-6)}$\Msun.
For $v_{\rm bc}=0$, the first peak is at around several hundred \Msun,
which is consistent with a large suite of radiative hydrodynamical
simulations \citep{Hirano2015}, considering only the H$_2$ mode.  In
their simulations, \p\ stars with mass less massive than 100\Msun\ form;
however, the formation mode of majority of them is via both H$_2$ and
HD. As described in~\ref{sec:model:pop3_imf}, our model takes the HD
mode into account.  However, the critical mass of the HD mode is
always smaller than the \p\ critical mass except for $z \lesssim 15$
as shown in Figure~\ref{fig:mcrit}.  
In fact, a large fraction of less massive \p\ stars form at $10 \lesssim z \lesssim 15$ 
in \citet{Hirano2015}.  After this redshift, the LW background grows
sufficiently in our model, 
and therefore, no less massive \p\ stars
are born via HD cooling even in the $v_{\rm bc}=0$ case.  On the other
hand, \citet{Hirano2015} adopted a smaller simulation box (3\hMpc)
than our study (16\hMpc), possibly resulting in weaker LW feedback
and the formation of less massive \p\ stars.

The second peak corresponds to the atomic cooling halos, where
supermassive stars are born.  Such second peak is also seen in a
semi-analytic study by \citet{Toyouchi2023} for a mass distribution of
primordial stars in a high-$z$ progenitor halo of Quasi-Stellar Object
(QSO).  As shown in Figure~\ref{fig:mcrit}, the critical halo mass
increases with the LW flux, and reaches the atomic cooling limit at
$z\sim20$ when the LW flux is sufficiently high.  The critical mass of
the atomic cooling halo does not depend on the value of streaming
velocity in our model, hence the \p\ mass formed in atomic
cooling halos tends to be similar regardless of the value of streaming
velocity.  The second peak is higher with the streaming velocity
because the existence of the streaming velocity increases the critical
halo mass; therefore, there are more pristine massive halos at $z\sim20$ in
the higher velocity model than in lower one.

The right panel of Figure~\ref{fig:imf} is another look of the \p\ mass
distribution, the cumulative probability of the number of \p\ stars as a
function of \p\ mass.  The sharp rise at $2 \times
10^5$\Msun\ corresponds to the atomic cooling halos.  The fractions of
such supermassive stars are about 2, 21, 57, and 83\% for 
$v_{\rm bc}=0$, 1\sv, 2\sv, and 3\sv, respectively. 
These results indicate that the majority of
\p\ stars are supermassive in high streaming velocity regions.

\begin{figure*}
\centering
\includegraphics[width=0.9\linewidth]{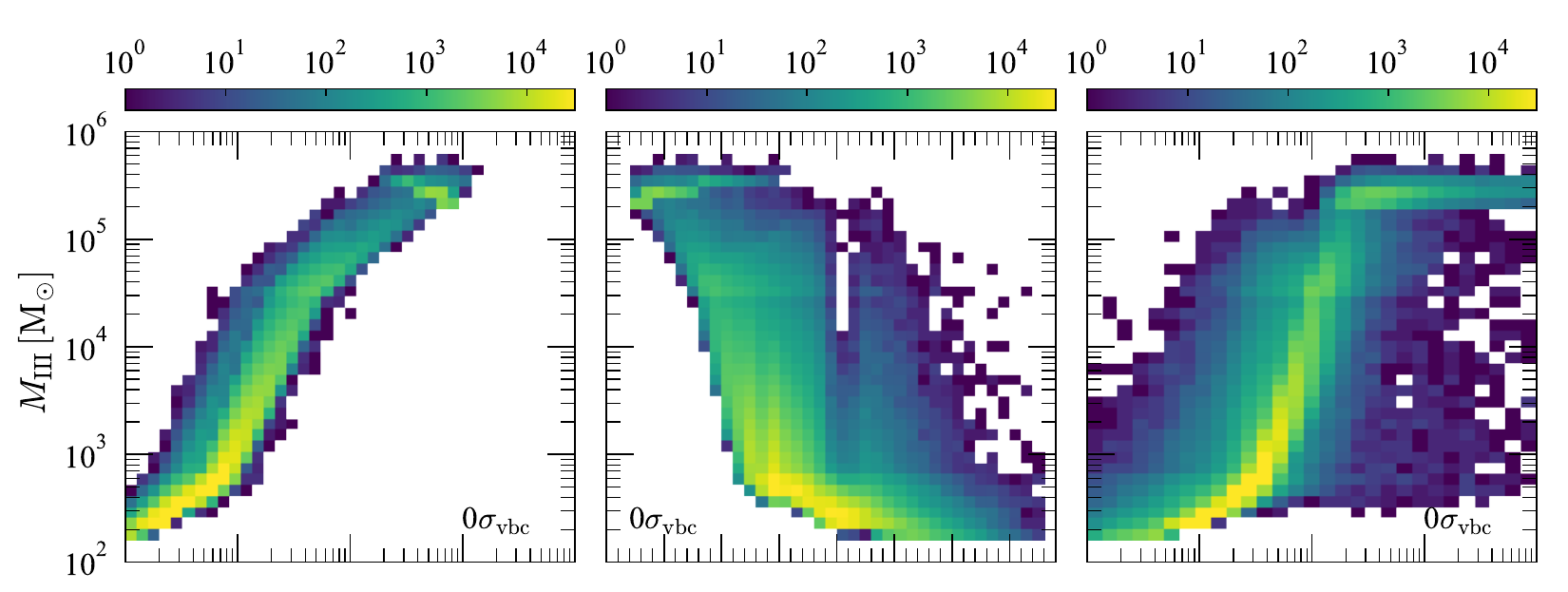}
\includegraphics[width=0.9\linewidth]{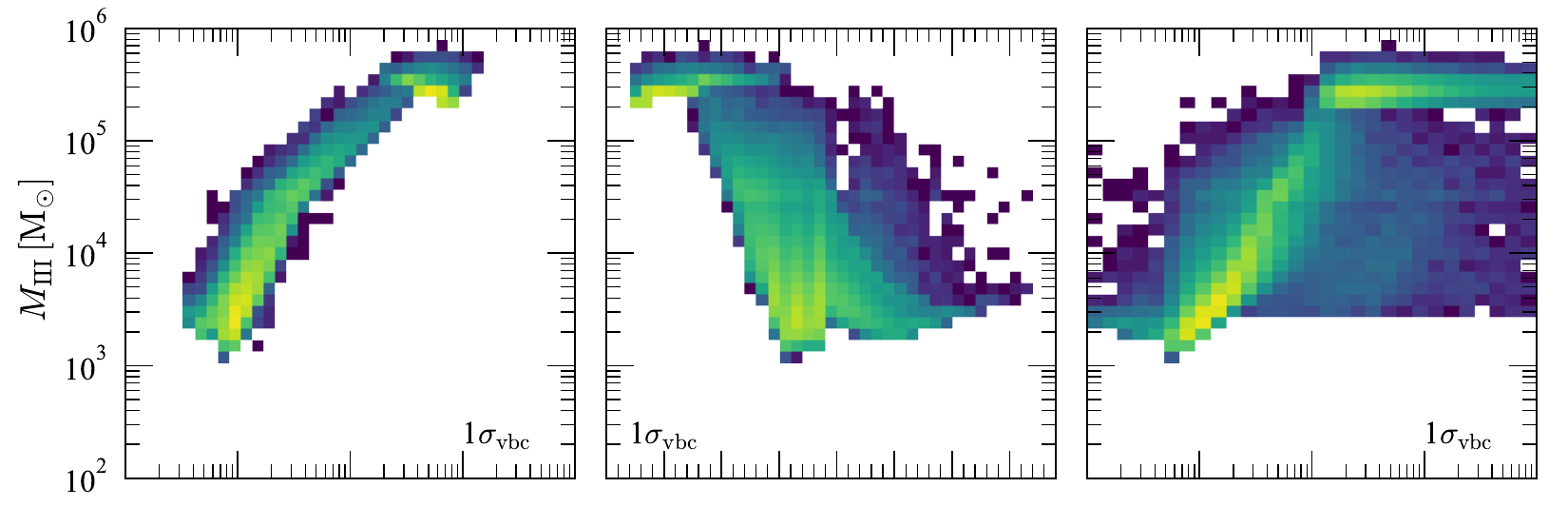}
\includegraphics[width=0.9\linewidth]{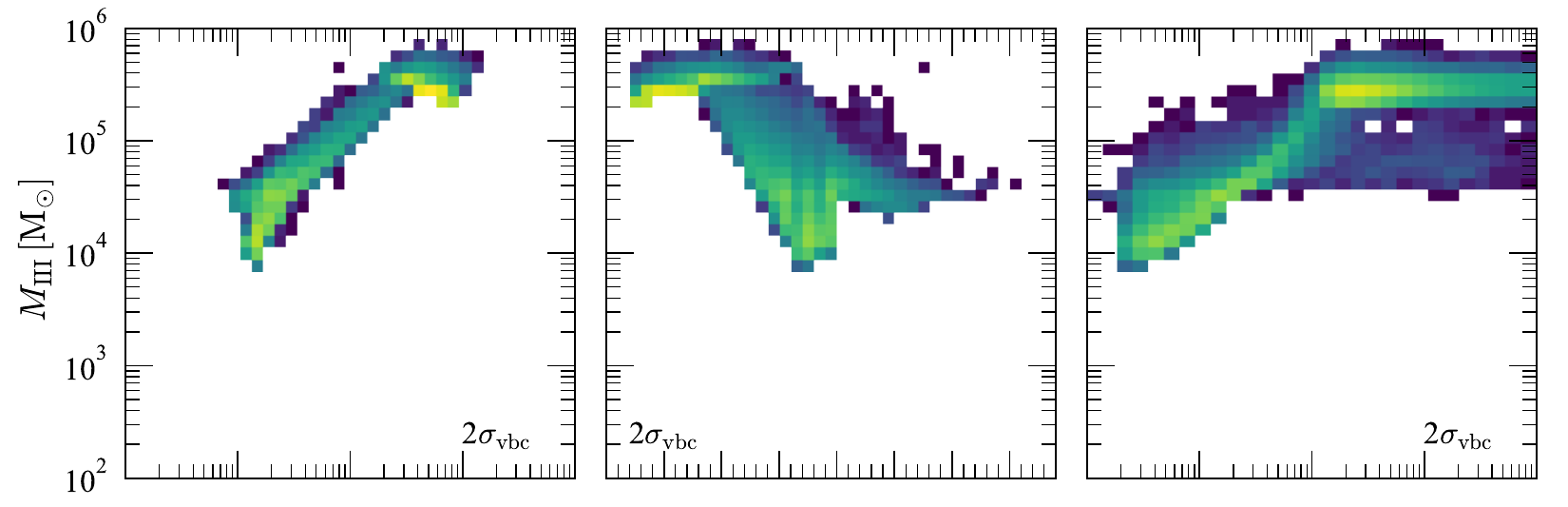}
\includegraphics[width=0.9\linewidth]{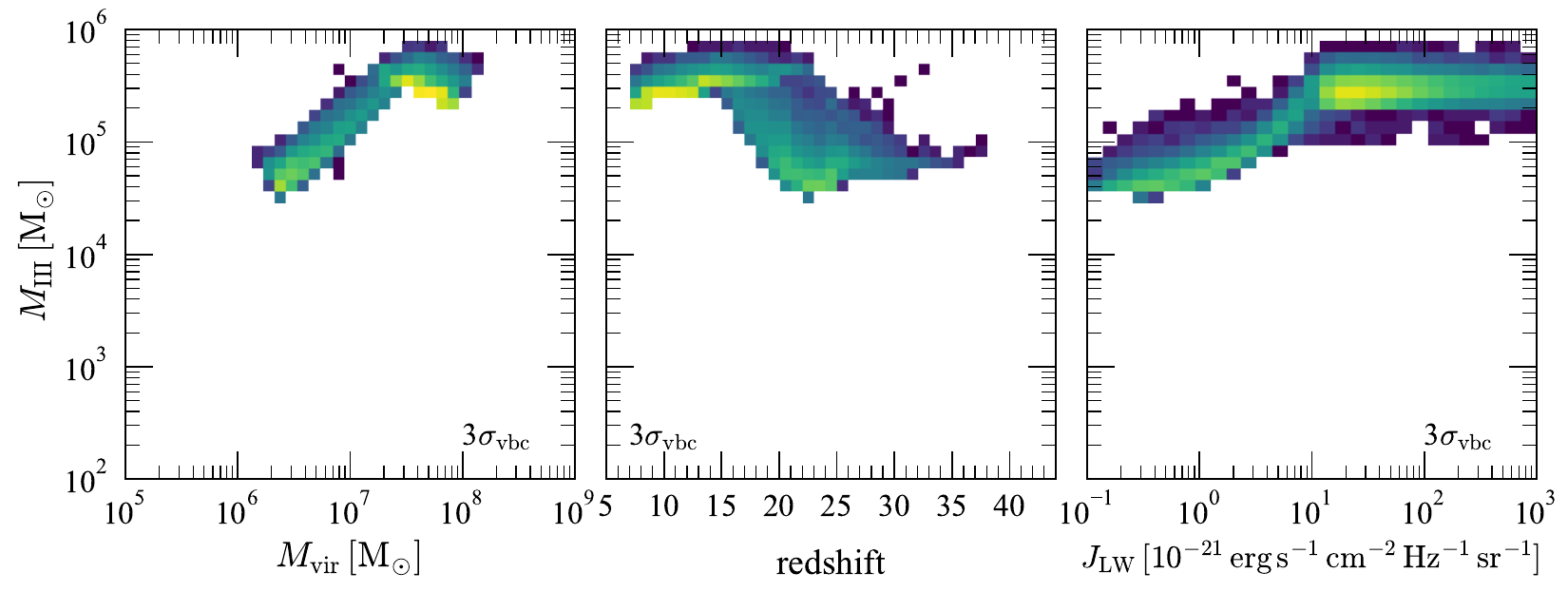}
\caption{
Dependence of \p\ mass on virial mass (left), redshift (middle), and
Lyman-Werner flux (right) of host halos.  From top to bottom, model results with
$v_{\rm bc}=0,1,2$, and 3\sv\ are presented.
Color bars show the number of \p\ stars at each pixel.
}
\label{fig:cor}
\end{figure*}

Figure~\ref{fig:cor} demonstrates the dependence of \p\ mass on
properties of halos: virial mass, redshift, and LW flux of host halos.
All panels exhibit two sequences.  In the panel of the virial mass,
the number of \p\ has peaks around the lowest \p\ mass. Regardless of the
value of streaming velocity, the corresponding redshift and LW flux
are around $20 \lesssim z \lesssim 25$ and $J_{\rm LW} \lesssim 10$,
respectively.  The lowest mass increases with the value of streaming
velocity as already seen in Figure~\ref{fig:imf} because the typical
halo mass also increases with the value of streaming velocity. The
\p\ mass gradually increases with the virial mass because the mass
accretion rate also increases.  The number density of \p\ mass
gradually decreases with the virial mass, but finally reaches second
peak, which corresponds to the atomic cooling limit.  Because the
critical mass of atomic cooling does not depend on the value of
streaming velocity in our model, the corresponding virial mass and
\p\ mass do not change across the value of streaming velocity, 
while the number density is higher in the second peak than in the first one 
for 2\sv\ and 3\sv\ models
as already demonstrated in Figure~\ref{fig:imf}.

The second peaks exist around $7.5 \lesssim z \lesssim 15$.  As the
value of streaming velocity increases, the second peaks emerge from
slightly higher redshift and the number density therein also becomes
higher.  Because the H$_2$ critical mass also increases with
the value of streaming velocity, more halos can enter the atomic
cooling limit in higher velocity models, as shown in
Figure~\ref{fig:mcrit}.  This is also the reason that the LW flux at
the second peak is slightly more broadly distributed in higher
velocity models toward the lower value.  A large fraction of halos can
cool via H$_2$ in lower velocity models, before the LW flux grows
sufficiently.  Therefore, they have a smaller number of atomic cooling
halos with low LW flux.

%%%%%%%%%%%%%%%%%%%%%%%%%%%%%%%%%%%%%%%%%%%%%%%%%%%%%%%%%%%%%%%%%%%%%%%%%%%%%%
%%%%%%%%%%%%%%%%%%%%%%%%%%%%%%%%%%%%%%%%%%%%%%%%%%%%%%%%%%%%%%%%%%%%%%%%%%%%%%

\subsection{Star formation rate density}\label{sec:sfrd}

\begin{figure*}
\centering \includegraphics[width=0.49\linewidth]{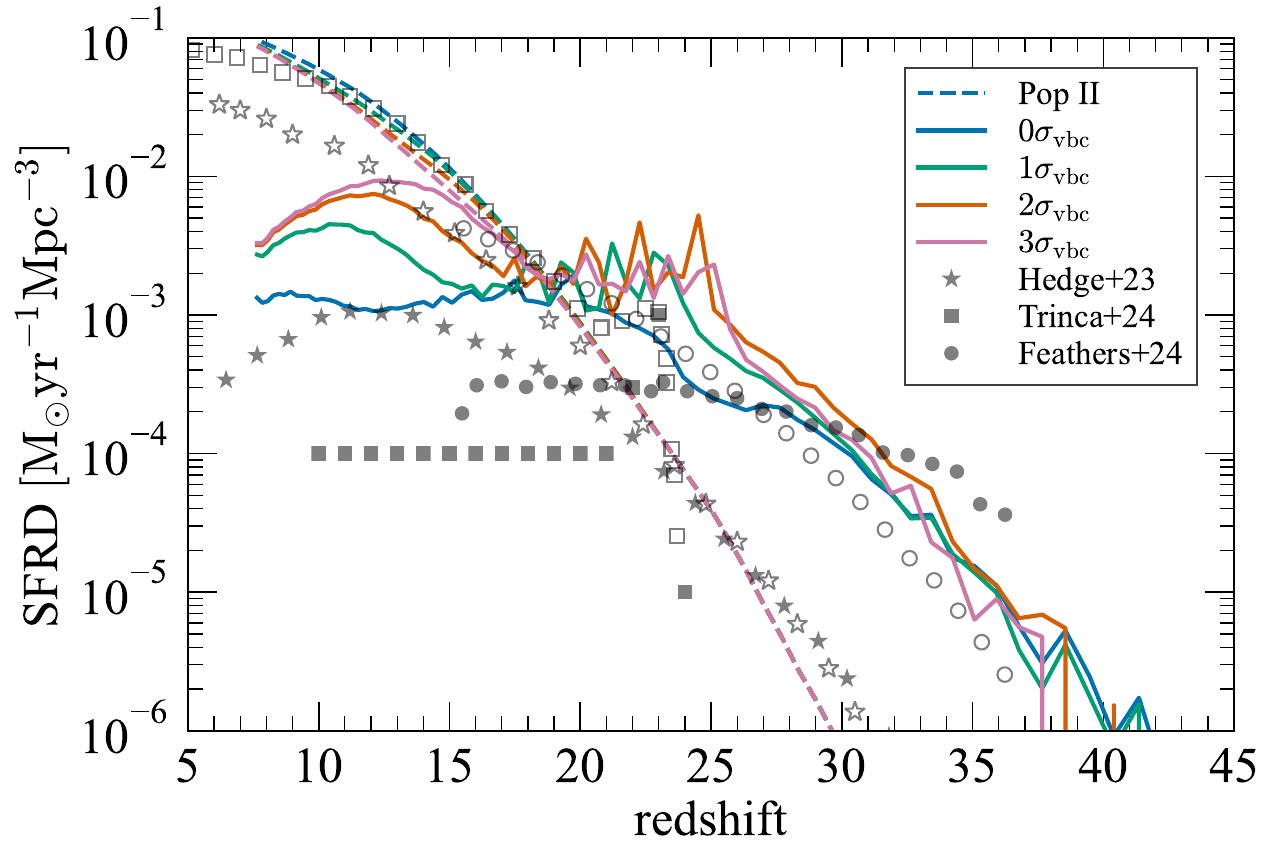}
\includegraphics[width=0.49\linewidth]{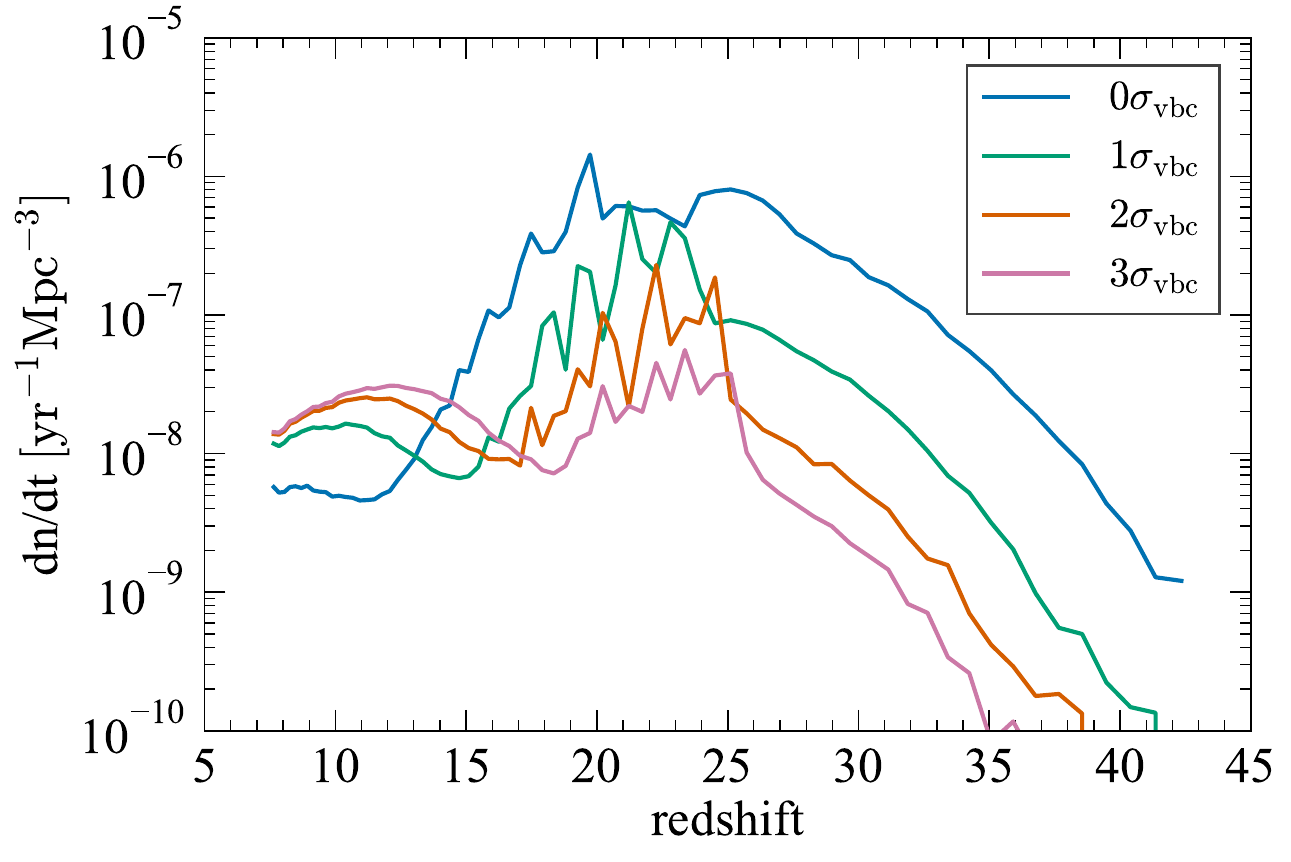}
\caption{
(Left) Star formation rate density (SFRD) of \p\ (solid curves) and \pii\ (dashed) stars. Four
model results with the different values of streaming velocity, $v_{\rm bc}=0$,
1\sv, 2\sv, and 3\sv, are shown.  
Filled and open symbols show other semi-analytic results for \p\ and \pii, respectively \citep{Hedge2023, Feathers2024,Trinca2024}. 
(Right) Formation rate density of \p-forming halos  per \Mpccube and year.
}
\label{fig:sfrd}
\end{figure*}

\begin{figure*}
\centering
\includegraphics[width=0.49\linewidth]{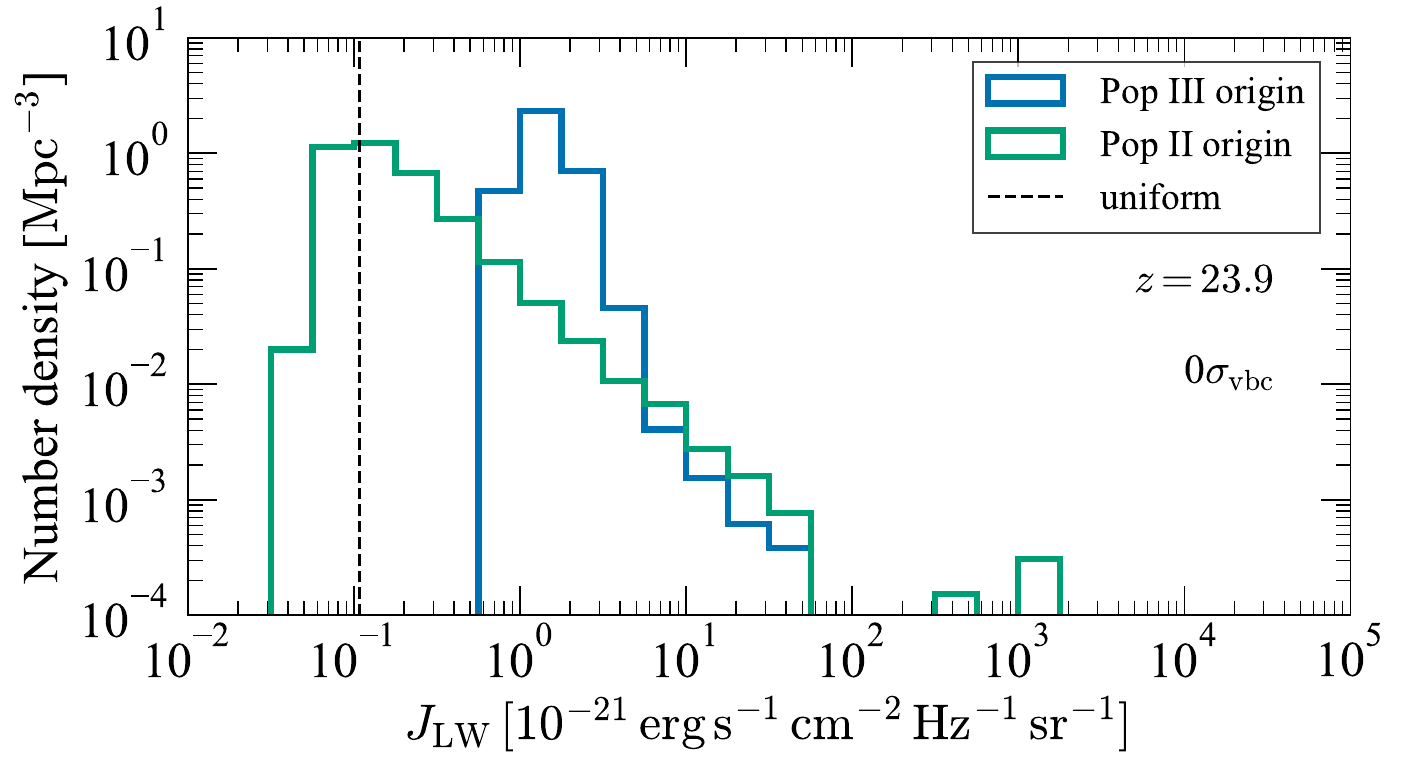}
\includegraphics[width=0.49\linewidth]{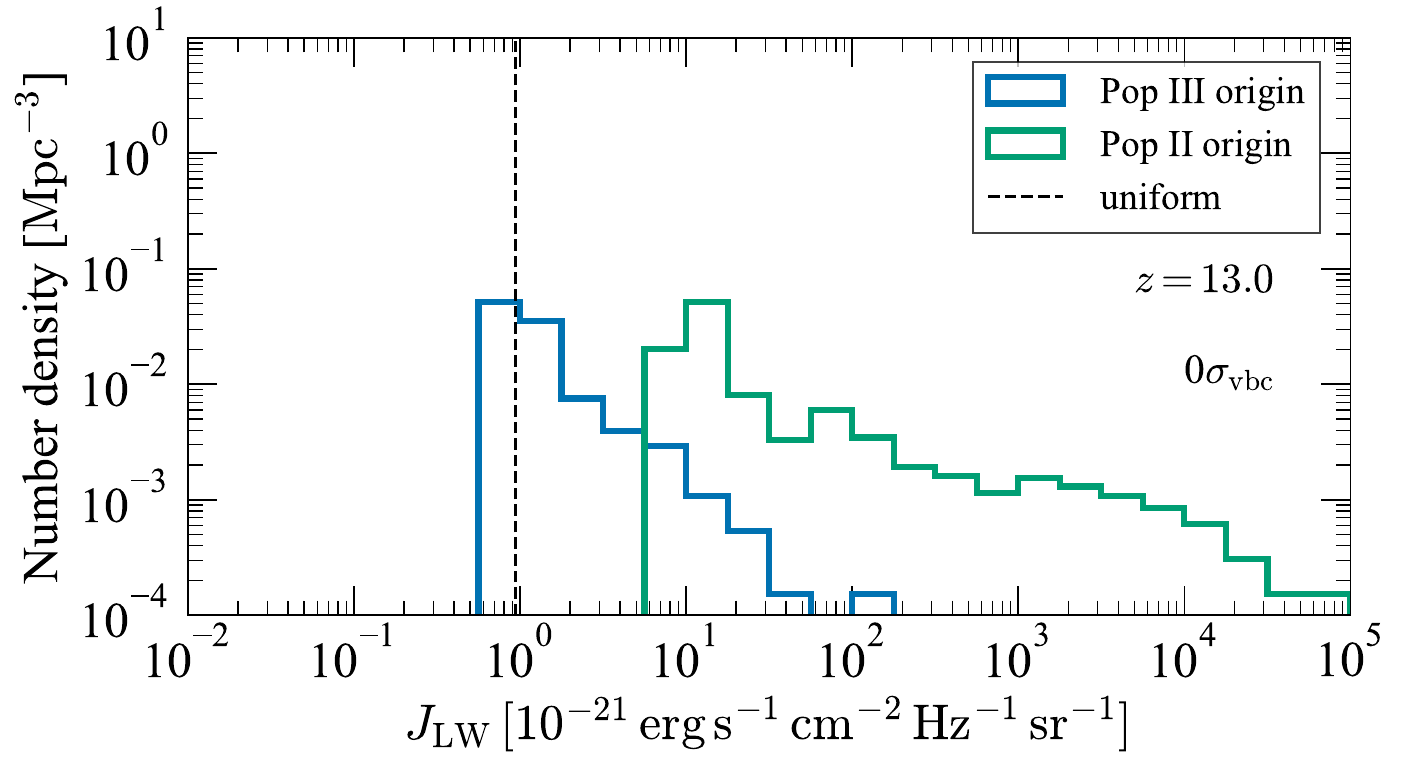}
\caption{ 
Number density of \p\ stars as a function of the Lyman-Werner flux
radiated on their halos at $z=23.9$ (left) and $z=13.0$ (right) for
$v_{\rm bc}=0$ model.  Histograms show \jlw\ from \p\ and \pii\ sources
in the fiducial (self-consistent) model.  We divide a single-digit
\jlw\ range into four logarithmic bins and calculate the number
density in each bin.  Vertical dashed lines show results from the
spatially uniform \jlw\ model.
}
\label{fig:jlw}
\end{figure*}

Figure~\ref{fig:sfrd} shows the star formation rate density (SFRD) of
\pandii\ stars and the formation rate density of \p-forming halos as a
function of redshift for four models.  The \p\ formation begins around
$z\sim40$, and the rate rises with evolving redshift and reaches the first
peak at $20 \lesssim z \lesssim 25$.  Around this redshift, LW flux
sufficiently grows to prevent subsequent \p\ formation
as shown in Figure~\ref{fig:jlw}, 
which is the distribution of the LW flux radiated on each \p-forming halo 
(see also Figure~\ref{fig:visual}).
Hence, the formation rate stagnates and falls regardless of the model. Then, the formation rate rises again as new supermassive \p\ stars are born in
atomic cooling halos. As already seen in previous Figures, the
\p\ formation rate reaches the second peak around $7.5 \lesssim z \lesssim
15$, and the peak is on earlier redshift with increasing streaming velocity.
In the case of SFRD, it stagnates but does not fall once it reaches
the first peak. However, the second peak is also observed as well as the
formation rate.  These double peaks have not been reported in any
literature probably because insufficient computational volume was used
\citep[e.g.,][]{Visbal2020} or spatially uniform LW background was
assumed \citep[e.g.,][]{Hartwig2022}.
In fact, a large fraction of \p-forming halos shows high LW flux
($J_{\rm LW} > 100$ at $z=13$) in our self-consistent model as shown
in Figure~\ref{fig:jlw}, whereas $J_{\rm LW} \sim 1$ in the uniform
model.  In \S~\ref{sec:lw_model}, we further scrutinize the impact of
the LW feedback model on the statistics

Figure~\ref{fig:sfrd} also shows other semi-analytic results
\citep{Hedge2023, Feathers2024,Trinca2024}.  There is large diversity
of \p\ SFRD, and nearly one to two orders of difference is observed at
$z<25$, reflecting the difference of adopted physical models such as
the critical halo mass, \p\ mass, and feedback.  On the other hand, the
\pii\ SFRD agrees well between results by our model,
\citet{Hedge2023}, and \citet{Trinca2024}.

Figure~\ref{fig:jlw} shows that the LW flux is dominated by
\p\ sources at earlier epoch and \pii\ sources at later epoch,
corresponding to the SFRD. As the \pii\ SFRD increases, the flux from
\pii\ increases and the contribution from \p\ becomes sub-dominate, as
also indicated by Figure ~\ref{fig:visual}.  Supermassive stars are
born close to where the \pii\ stars are concentrated and the LW flux
from them is also high.  Those pictures and the distribution of the LW
flux are broadly consistent with the previous studies
\citep{Agarwal2012, Chon2016}.
There is difference in the minimum value of flux from \p\ sources
probably due to the adopted \p\ mass model. 

The \p\ SFRD is higher with increasing streaming velocity, while the
formation rate density of \p-forming halos shows the opposite 
trend in the H$_2$ cooling regime. These results can be understood as
follows.  The typical \p-forming halo mass increases with the value
of streaming velocity; hence, the formation rate of such halos is lower
due to the absence of massive halos. In contrast, the \p\ mass
increases monotonically with halo mass as already shown in the
previous section. The combination of these effects is the cause of this
opposite trend. In contrast, in the atomic cooling regime, both the
SFRD and \p-forming halo formation rate are higher with increasing
streaming velocity.

In the \pii\ SFRD, we observe no significant difference at $z \gtrsim
20$ and a slight difference after this redshift between the four
models. Throughout the models, the \pii\ formation begins on halos
that exceed $M_{\rm th}=10^7$\Msun\ and are enriched by other stars.  By
$z\sim20$, these \pii-forming halos are easily enriched by \p\ stars
regardless of the value of streaming velocity because the \p\ critical
mass is always less than $M_{\rm th}$ until this redshift, 
as seen in Figure~\ref{fig:mcrit} 
(\mcrit$ \lesssim 10^6$\Msun\ at $z=20$ and \jlw$=0$
for $v_{\rm bc}=0$ and 1\sv) and nearby halos  should be
massive enough to host \p\ stars even for the high \sv\
case due to the nature of halo clustering.  
Even in the 3\sv\ and \jlw$=0$ case, 
\mcrit$\sim3\times10^6$\Msun\ is still less than $M_{\rm th}$.
Consequently, the global
\pii\ SFRD is similar across the four models.
After that redshift, given the high LW intensity, 
the critical mass rapidly increases with decreasing redshift 
and the value of streaming velocity, and then finally 
reaches the atomic cooling mass, which is higher than $M_{\rm th}$. 
As a result, the contribution to the \pii\ SFRD from 
halos with the mass around $M_{\rm th}$ formed at a late time
becomes smaller with increasing the value of streaming velocity. 
This is the reason why the the \pii\ SFRD is slightly smaller 
as the value of streaming velocity increases.

%%%%%%%%%%%%%%%%%%%%%%%%%%%%%%%%%%%%%%%%%%%%%%%%%%%%%%%%%%%%%%%%%%%%%%%%%%%%%%
%%%%%%%%%%%%%%%%%%%%%%%%%%%%%%%%%%%%%%%%%%%%%%%%%%%%%%%%%%%%%%%%%%%%%%%%%%%%%%
\subsection{Box size effect}\label{sec:box_convergence}

\begin{figure*}
\centering
\includegraphics[width=0.49\linewidth]{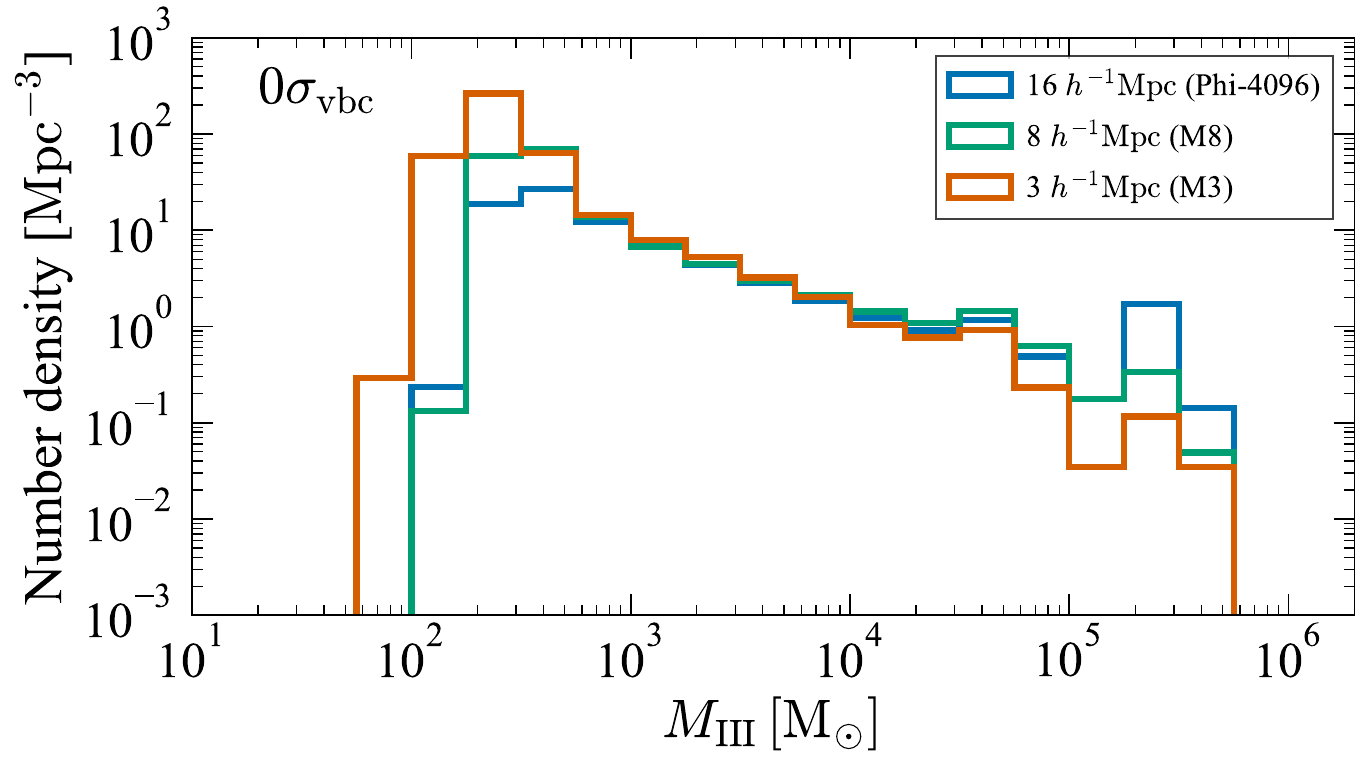}
\includegraphics[width=0.49\linewidth]{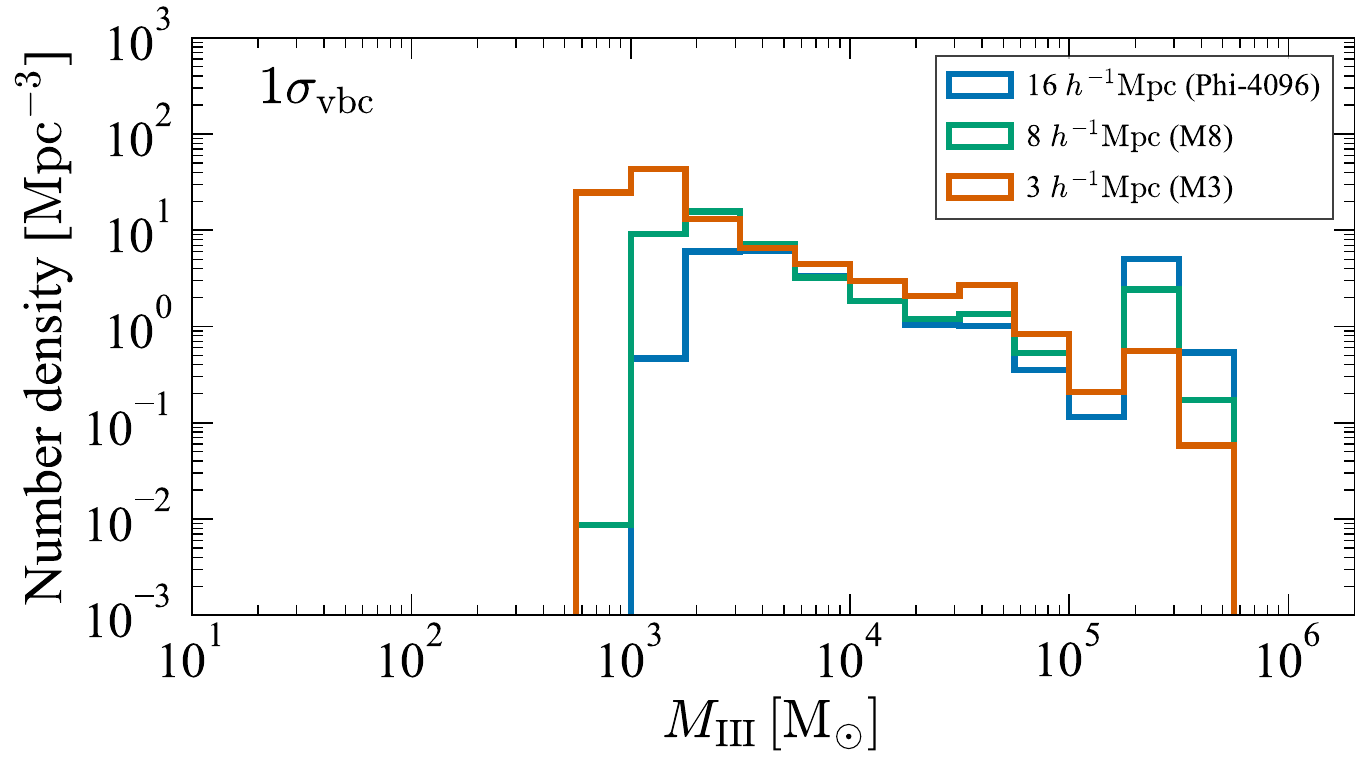}
\includegraphics[width=0.49\linewidth]{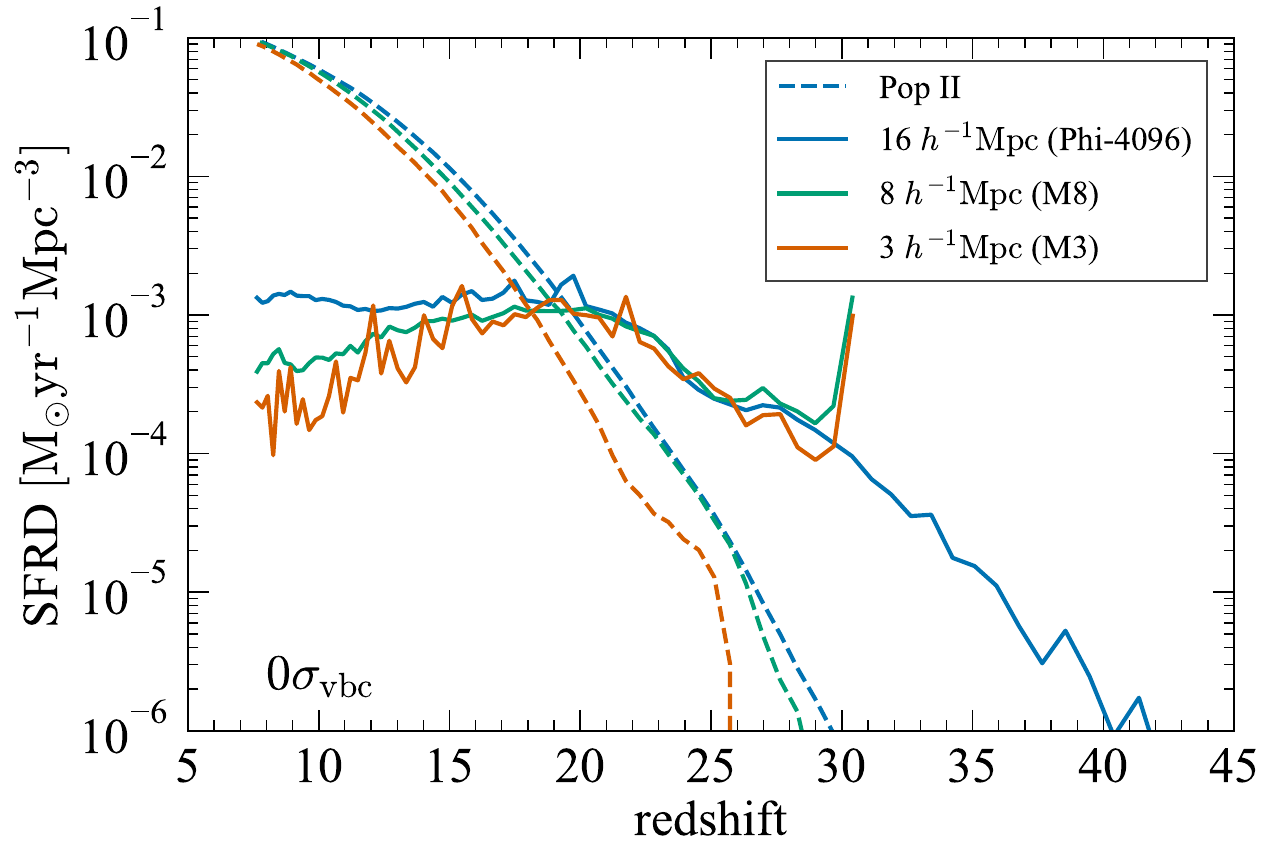}
\includegraphics[width=0.49\linewidth]{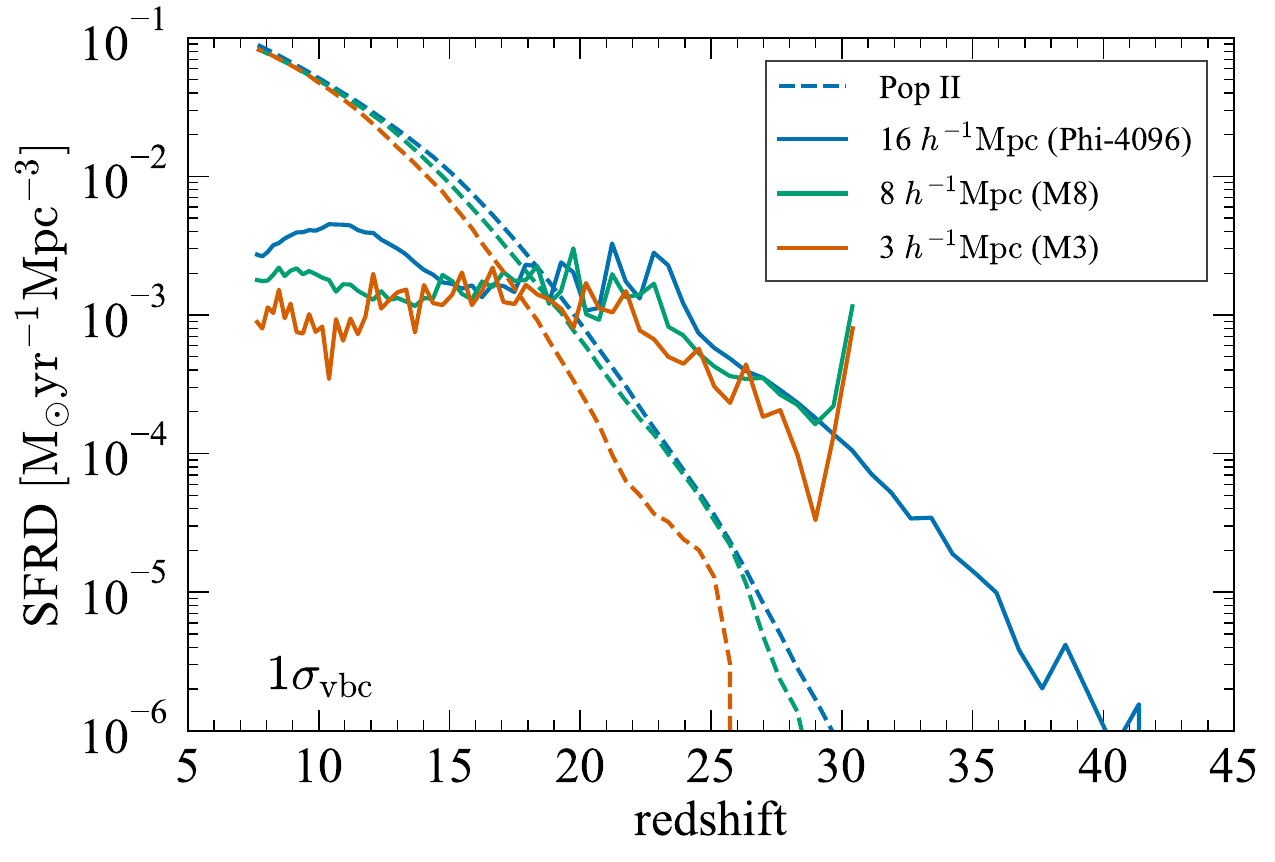}
\includegraphics[width=0.49\linewidth]{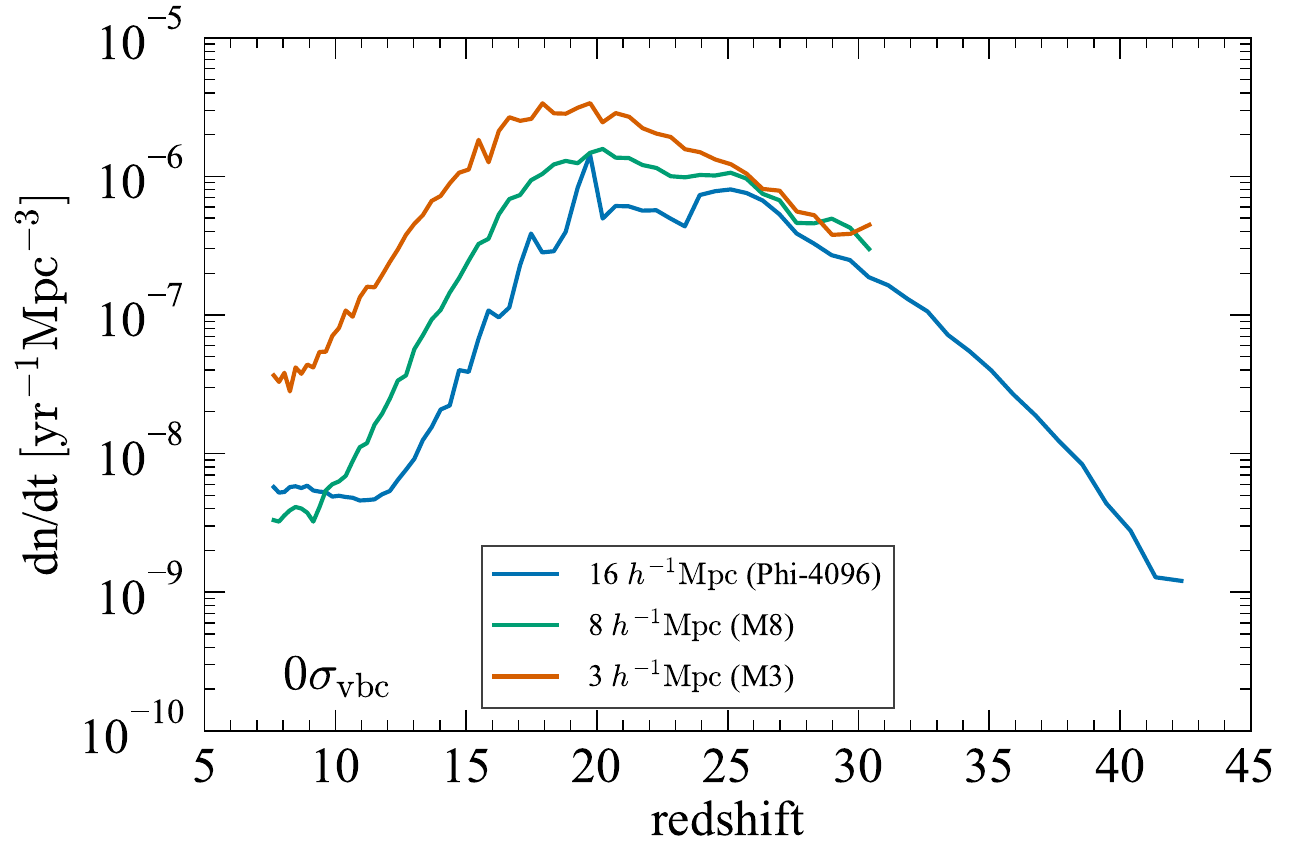}
\includegraphics[width=0.49\linewidth]{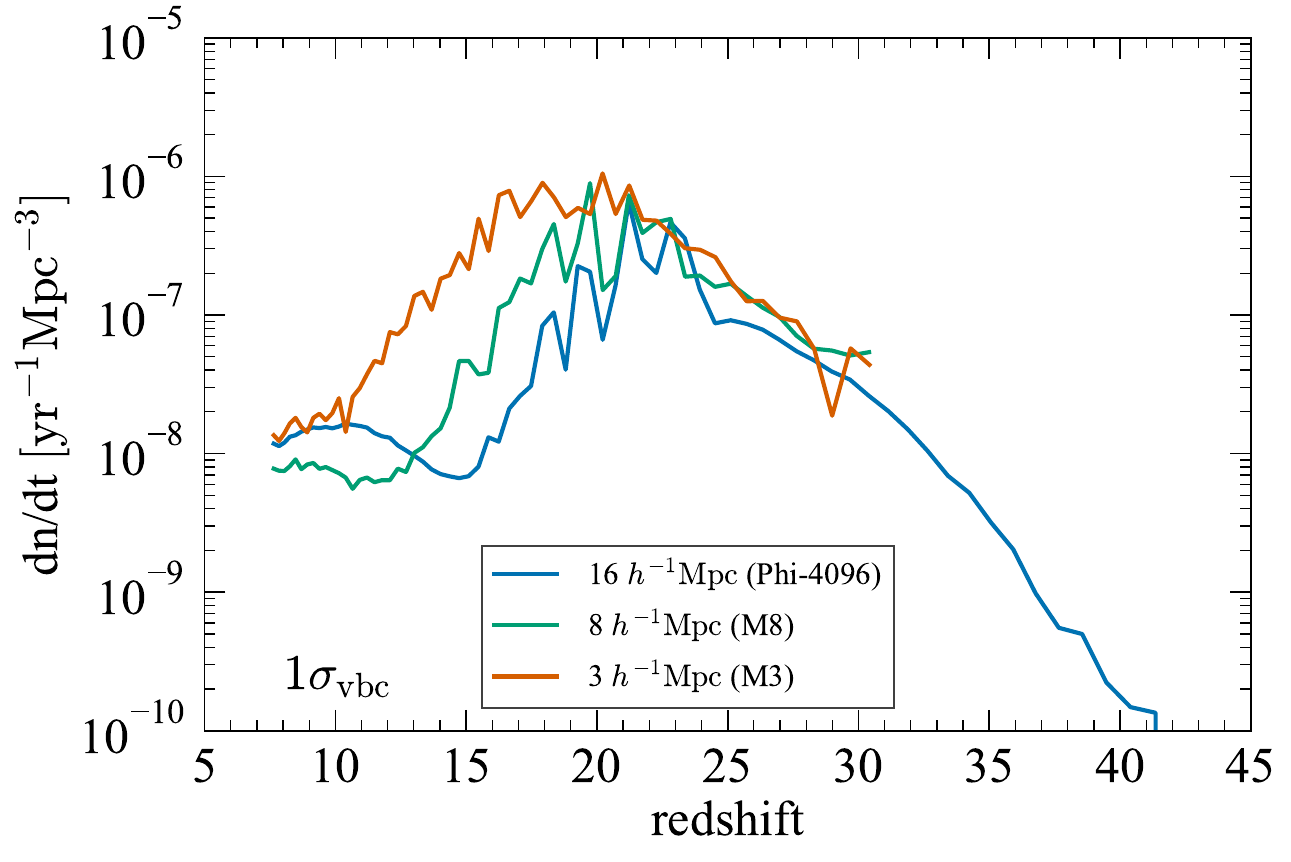}
\caption{
Comparison of model results calculated on simulations with different box sizes
(16, 8, and 3 \hMpc\ for \phif, \Me, and \Mt, respectively).
Left and right three panels show results of the model with 
$v_{\rm bc} = 0$ and 1\sv, respectively. 
(Top) Number density of \p\ stars formed at all redshift as a function of the
\p\ mass in a halo, same as Figure~\ref{fig:imf}.  (Middle) Star
formation rate density, same as Figure~\ref{fig:sfrd}.  (Bottom)
Formation rate density of \p-forming halos, same as
Figure~\ref{fig:sfrd}.  
}
\label{fig:box_conv}
\end{figure*}

In this section, we examine the effect of the simulation box size on
the \p\ IMF formed at all redshift and SFRD.
Figure~\ref{fig:box_conv} shows the comparison of model results calculated on
simulations with different box sizes (\phif, \Me, and \Mt), which are
16, 8, and 3\hMpc.  The results with $v_{\rm bc} = 0$ and
1\sv\ are shown in the left and right panels,
respectively.  In the \Mt\ box, we observe considerably smaller number density
of supermassive stars than the \phif\ run for both $v_{\rm bc} = 0$
and 1\sv\ cases.  These are highlighted in the \p\ SFRD
plots, not showing the clear second peak for the \Mt.  The \pii\ SFRD
of the \Mt\ box shows a delay from the others due to the absence of
large-scale fluctuations, resulting in insufficient LW feedback from
\pii\ stars and the suppression of the formation of atomic cooling halos.
Correspondingly, the number density of \p-forming halos  increases with
decreasing box size due to insufficient LW feedback, as shown in
the formation rate plots.

As the box size increases, the number density of supermassive stars increases.
In the \Me\ box, we observe smaller number density of supermassive stars
than the \phif\ run for the $v_{\rm bc} = 0$ case, while it
approaches asymptotically to the result of \phif\ run for the 
$v_{\rm bc} =1\sigma_{\rm vbc}$ case.  
Those are also highlighted in the \p\ SFRD
and the formation rate plots where the second peak 
in the \Me\ is weak for $v_{\rm bc} = 0$ 
and appears for $v_{\rm bc} = 1\sigma_{\rm vbc}$.

In summary, at least a 8\hMpc\ box is necessary to capture
the atomic cooling regime in the case with the streaming velocity,
while even a larger box is necessary in the case without the streaming
velocity. This might be one of the reasons why the second peak was not observed
in previous semi-analytic studies that used much smaller boxes (e.g.,
2\hMpc\ for \citealt{Visbal2020}).

%%%%%%%%%%%%%%%%%%%%%%%%%%%%%%%%%%%%%%%%%%%%%%%%%%%%%%%%%%%%%%%%%%%%%%%%%%%%%%
%%%%%%%%%%%%%%%%%%%%%%%%%%%%%%%%%%%%%%%%%%%%%%%%%%%%%%%%%%%%%%%%%%%%%%%%%%%%%%
\subsection{Impact of the spatial variation of LW feedback}\label{sec:lw_model}

Many previous studies adopted spatially uniform LW background
\citep[e.g.,][]{Hartwig2022,Hedge2023,Bovill2024,Feathers2024},
while others took the spatial variation into account
\citep[e.g.,][]{Agarwal2012, Visbal2020, Ventura2024} within a few
\hMpc\ simulation box. However, it is essential to use a larger box to
investigate the effect of inhomogeneous LW background because spatial
inhomogeneities could dominate the intensity up to a distance of
$\sim$ 100 proper kpc \citep{Incatasciato2023}.  In this paper, we
first examine the impact of the spatial variation of LW feedback
within much larger spatial volume compared to the previous studies.

In our self-consistent model, the spatial variation of LW feedback
naturally emerges. We compare it with a model adopting spatially
uniform LW background. To compute the uniform background, we
follow the formulae in \citet{Greif2006},
\begin{eqnarray}
  \label{eq:greif2006}
  J_{\rm LW}(z) \simeq \frac{hc}{4 \pi m_{\rm H}} \eta_{\rm LW} \rho_{\star}(z) \qty(1+z)^3,
\end{eqnarray}
where, $h, c$, and $m_{\rm H}$ are the Planck constant, speed of
light, and hydrogen atom mass, respectively.  The constant
$\eta_{\rm LW}$ denotes the number of LW photon products per stellar
baryon; here we assume $\eta_{\rm LW} \simeq 10^4$ and $2 \times
10^4$ for \pandii\ stars, respectively.  We evaluate $\rho_{\star}(z)$
every step directly during model calculations, which is the comoving
mass density of living \p\ stars
and \pii\ stars formed within 5 Myr from the current substep as used in 
Eq~\eqref{eq:Johnson2013}.

\begin{figure}
\centering
\includegraphics[width=\linewidth]{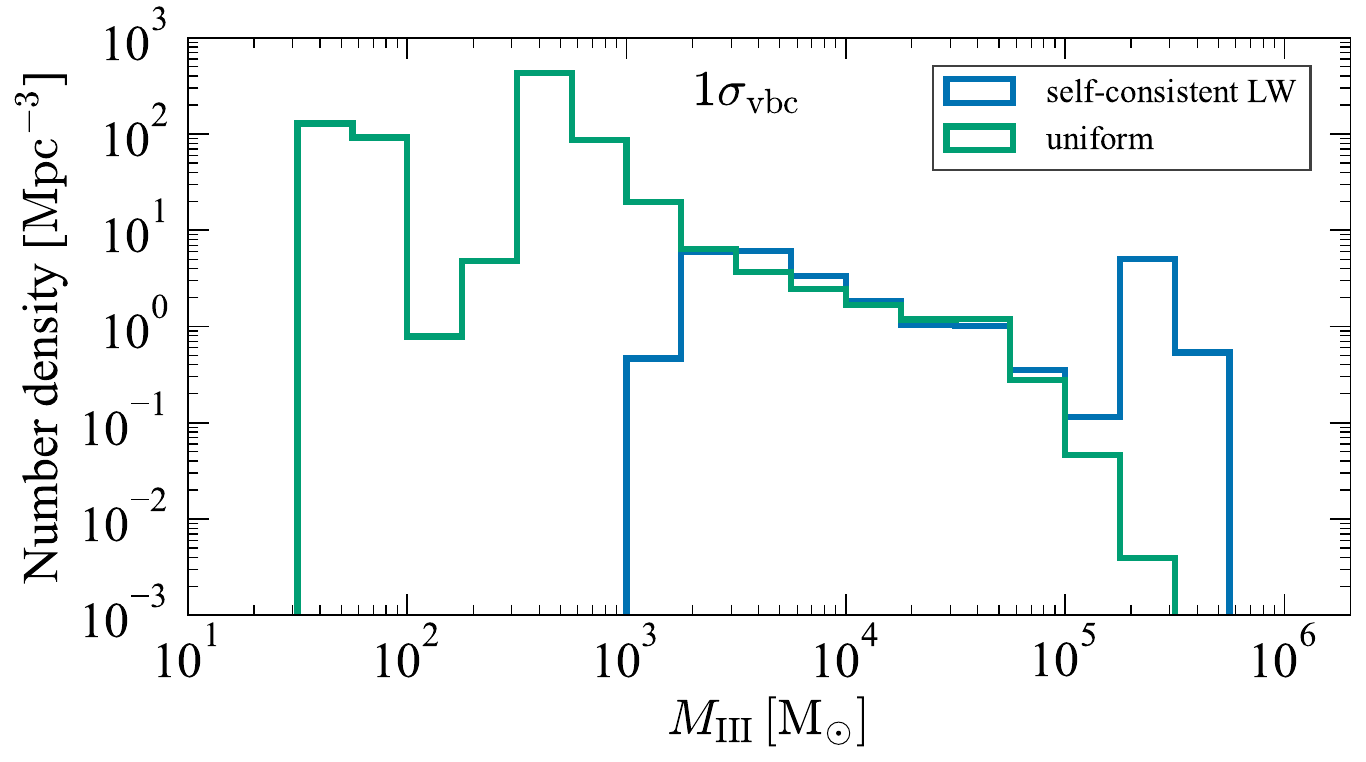}
\includegraphics[width=\linewidth]{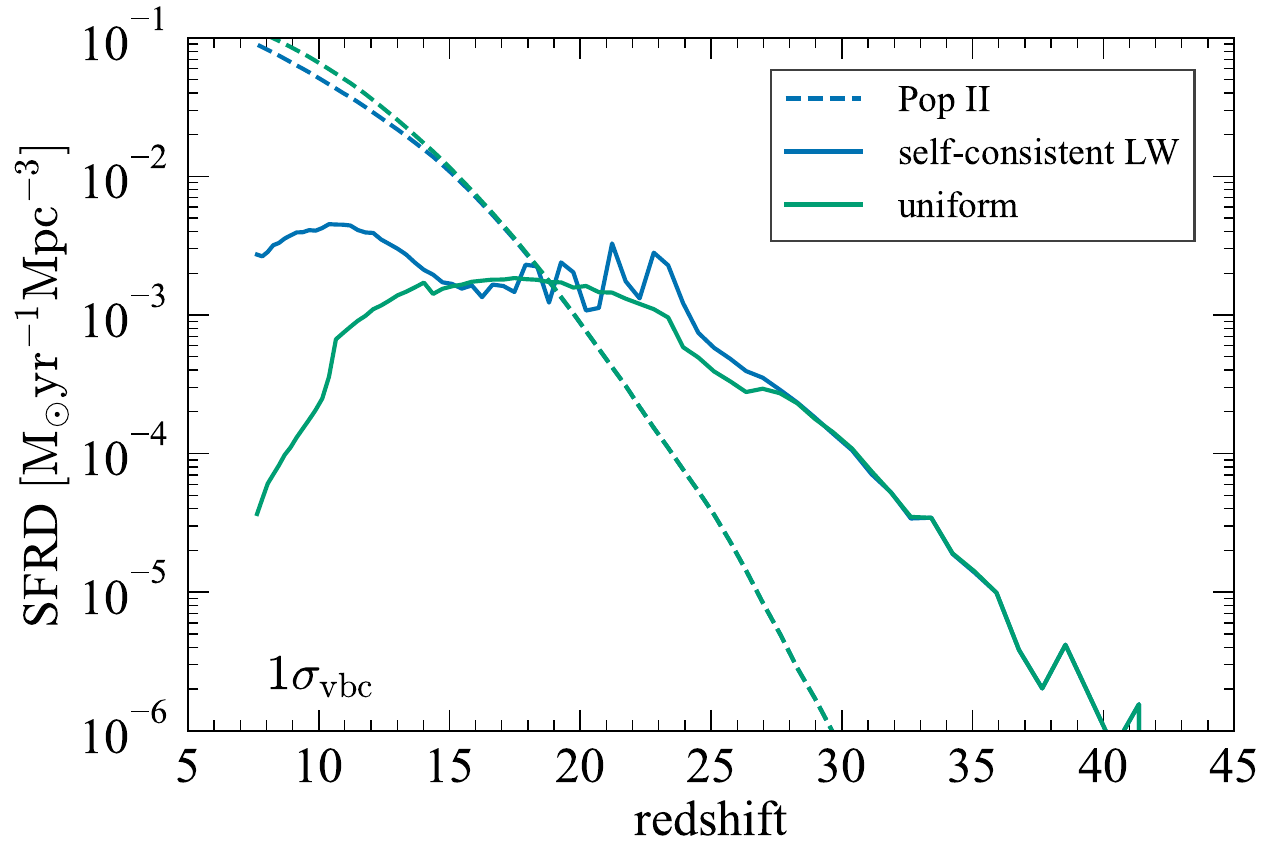}
\includegraphics[width=\linewidth]{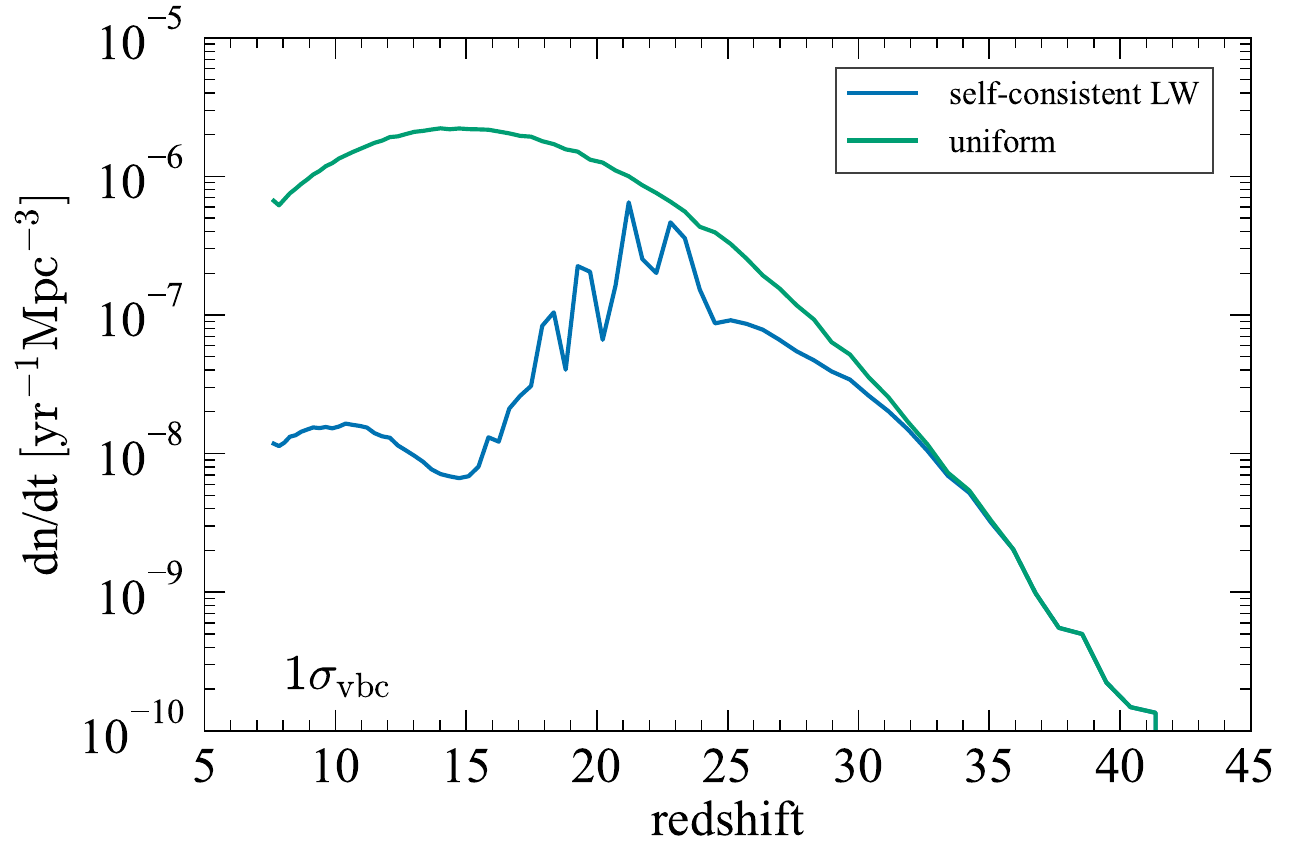}
\caption{
Comparison of different implementations of the LW feedback 
on the \phif\ simulation for the 
$v_{\rm bc} = 1\sigma_{\rm vbc}$ case. 
 Blue and green curves show results of the
fiducial (self-consistent) model and spatially uniform \jlw\ model, respectively.
(Top) Number density of \p\ stars formed at all redshift as a function of the
\p\ mass in a halo, same as Figure~\ref{fig:imf}.  (Middle) Star
formation rate density, same as Figure~\ref{fig:sfrd}.  (Bottom)
Formation rate density of \p-forming halos, same as
Figure~\ref{fig:sfrd}.  
}
\label{fig:lw_comp}
\end{figure}

Figure~\ref{fig:lw_comp} shows the comparison of self-consistent and
spatially uniform LW models on the \phif\ simulation for the $v_{\rm bc} =
1\sigma_{\rm vbc}$ case.  Both models give a similar formation rate
of halos  at $z \gtrsim 30$ (bottom panel), while it changes
dramatically after that, which corresponds to the epoch that
\pii\ SFRD starts to rise.  At that time, the {\it spatially average}
LW background does not grow sufficiently 
(see also Figure~\ref{fig:visual} and~\ref{fig:jlw}), 
therefore, the formation rate is
not suppressed in the uniform model. On the other hand, the
\p\ formation in halos  near to \pii-forming halos starts to be
suppressed effectively due to the {\it locally strong} LW background
in the self-consistent model.  This suppression gradually becomes
strong and the formation rate is reduced by a factor of $\sim10$ and
$100$ at $z\sim20$ and 15, respectively.  The second peak emerges in
the self-consistent model at $z<15$.  This is not observed in the
uniform model although the average formation rate is gradually
suppressed due to a well-grown uniform LW background at $z<15$.  As already
seen in section~\ref{sec:box_convergence}, at least a 8\hMpc\ box
is necessary to correctly capture the atomic cooling regime in the
case with the streaming velocity. Our results shown here indicate that
the self-consistent model is also necessary to capture it.

Intriguingly, the SFRD of \p\ stars is similar between both models
before the second peak. This can be understood as follows. In the
uniform model, most of the halos  are not radiated by enough LW photons
from nearby sources to photodissociate H$_2$ molecules within them,
therefore, the typical halo mass (and also \p\ mass) is smaller
than that of the self-consistent model. In fact, the \p\ mass plot
shows that the uniform model gives a number of low-mass \p\ stars and
even the HD cooling mode ($M_{\rm III} \lesssim 10^2$\Msun) emerges, which
is not observed in the self-consistent model regardless of the value of
streaming velocity, as discussed in section~\ref{sec:imf}. The
difference of typical \p\ mass results in the similar \p\ SFRD at $z
\gtrsim 15$.

%%%%%%%%%%%%%%%%%%%%%%%%%%%%%%%%%%%%%%%%%%%%%%%%%%%%%%%%%%%%%%%%%%%%%%%%%%%%%%
%%%%%%%%%%%%%%%%%%%%%%%%%%%%%%%%%%%%%%%%%%%%%%%%%%%%%%%%%%%%%%%%%%%%%%%%%%%%%%
%%%%%%%%%%%%%%%%%%%%%%%%%%%%%%%%%%%%%%%%%%%%%%%%%%%%%%%%%%%%%%%%%%%%%%%%%%%%%%
%%%%%%%%%%%%%%%%%%%%%%%%%%%%%%%%%%%%%%%%%%%%%%%%%%%%%%%%%%%%%%%%%%%%%%%%%%%%%%
\section{Discussions}\label{sec:discussion}

\subsection{Implications for the formation of supermassive black holes at high redshift}

If the seed of black holes is massive enough in a halo, 
it can be a progenitor of high-z QSO \citep[e.g.,][]{Latif2022b,Latif2023}.
In this section, we discuss whether the supermassive \p\ stars can be a
promising candidate of the seed of supermassive black holes (SMBHs) by
comparing the number density of these stars with that of high-z QSOs. 
The global average number density of supermassive stars 
per comoving \Mpccube\ (\ndsms) 
predicted in the four models is shown in Table~\ref{tab:nsms}. 
Assuming each of 20, 72, 8, and 0\% of the whole volume is 
the spatial region with the streaming velocity of
$v_{\rm bc}=0$, 1\sv, 2\sv, and 3\sv, respectively, 
the number density of supermassive stars is predicted to be 
9.54 and 5.23 per comoving \Mpccube\
for $M_{\rm III} > 10^4$ and $10^5$\Msun, respectively. 
The existence of streaming velocity enhances the global number of 
supermassive stars by a factor of 1.6$\sim$2.6.

\begin{figure}
\centering
\includegraphics[width=\linewidth]{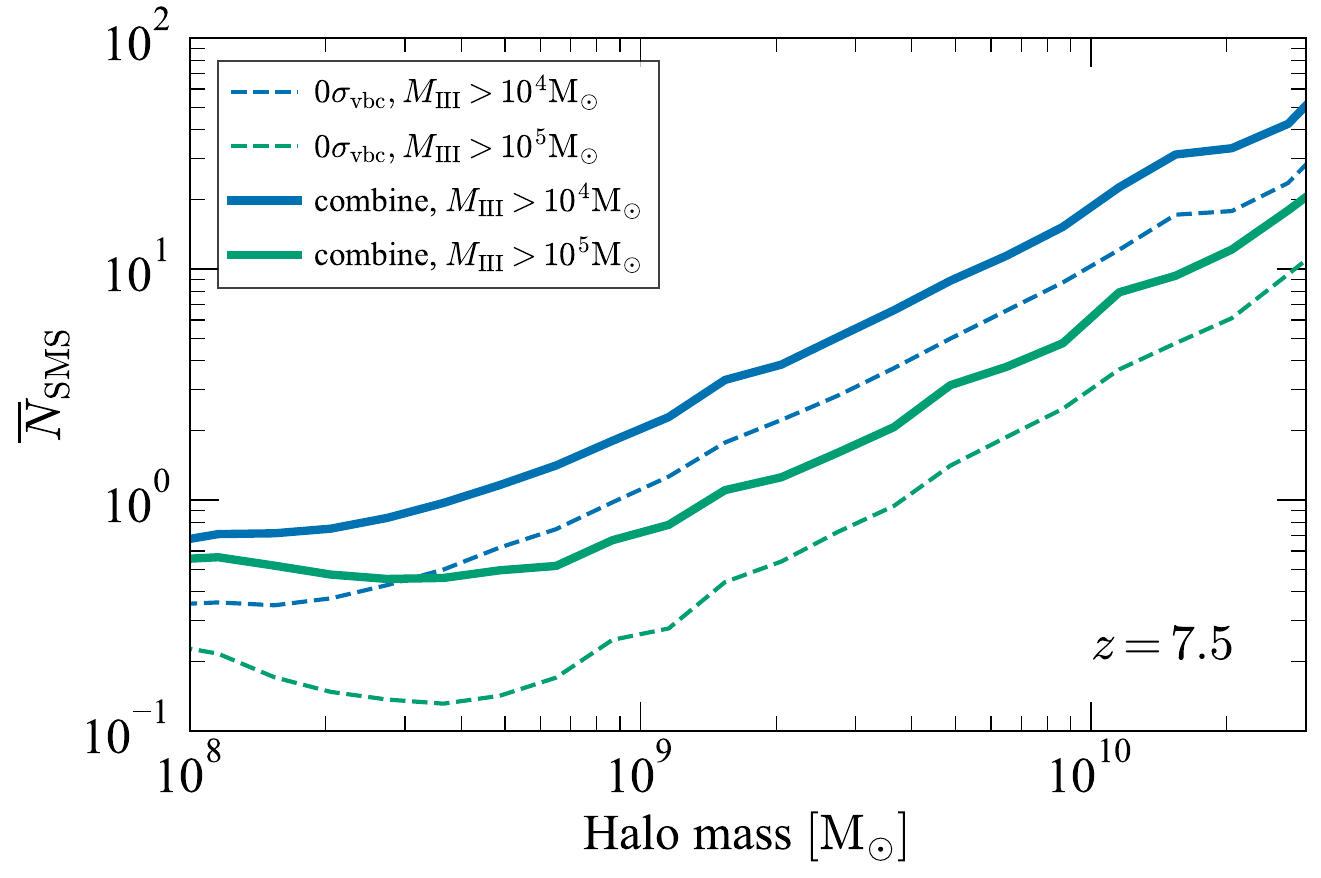}
\caption{
Average number of supermassive stars per halo (\nsms) 
as a function of halo virial mass at $z=7.5$. 
Blue and green curves show the results for 
$M_{\rm III} > 10^4$ and $10^5$\Msun, respectively. 
Thin and thick curves show the results for 
0\sv\ and the combination of four models, respectively
(details are provided in the text and Table~\ref{tab:nsms} caption).
}
\label{fig:nsms}
\end{figure}

The number density estimated above is the global average.  The high-z
QSOs should be hosted by massive halos at that era, therefore, we next
estimate the average number of supermassive stars per massive halo by
counting the number of progenitor halos where supermassive \p\ stars
with $M_{\rm III} > 10^4$ and $10^5$\Msun\ are born.
Figure~\ref{fig:nsms} shows the average number of supermassive stars
accreted by a halo (\nsms) as a function of halo virial mass at
$z=7.5$.  Without the streaming velocity, the number of supermassive
stars is unity in halos with $0.9\sim4\times 10^9$\Msun\ virial mass.
Intriguingly, this mass is quite similar with the prediction by
\citet{Chiaki2023}, which performed a similar semi-analytic
calculation on several regions from cosmological zoom-in simulations,
although the model details are quite different.  On the other hand,
taking the streaming velocity into account, the number of supermassive
stars is unity for $0.4\sim1 \times 10^9$\Msun, which is a factor of
2$\sim$4 less massive than the case without the streaming velocity.
Cosmological hydrodynamical simulations suggest that a halo with the
mass of several $10^9$\Msun\ and a BH seed with $10^{4-6}$\Msun\ can
produce a QSO at $z=6$ \citep{Bhowmick2022}, which is consistent with
our results.

\citet{Arita2023} estimated that the halo mass of QSOs at $z\sim6$ is
about $5.0 \times 10^{12}$\hMsun\ using clustering analysis of
spectroscopically identified QSOs. In our simulation, we do not have
halos with such mass at $z=7.5$ due to the limiting size of simulation
volume.  Extrapolating the result in Figure~\ref{fig:nsms}, a number
of supermassive stars could be hosted by such halos as halos merge
with each other into more massive halos.  Some of the stars could also
merge with each other to form SMBHs, and the others can orbit within a
main halo.  We do not consider mergers of stars (or BHs) in the
current model and plan to investigate this in future work.
Future observations is also essential to further constrain the \p\ formation 
scenario \citep[e.g., ][]{Shirakata2016, Matsuoka2023, Oogi2023}.

\begin{table}[t]
\centering
\caption
{Global average number density of supermassive stars per comoving \Mpccube\ 
  (\ndsms) for the four models.
  The integration results across the streaming velocity
  are shown in the bottom row (labelled ``combine''),
where each of 20, 72, 8, and 0\% of the whole volume is assumed to be
the spatial region with the streaming velocity of
$v_{\rm bc}=0$, 1\sv, 2\sv, and 3\sv, respectively. 
}
\label{tab:nsms}
\begin{tabular}{ccc}
\hline
\shortstack{\\ \sv \\ {}}&
\shortstack{\\ $n_{\rm SMS}$ \\ ($M_{\rm III} > 10^4$\Msun)} &
\shortstack{\\ $n_{\rm SMS}$ \\ ($M_{\rm III} > 10^5$\Msun)} \\
\hline 
0 & 5.85 & 2.04 \\
1 & 9.96 & 5.71 \\
2 & 14.93 & 8.87 \\
3 & 12.53 & 10.76 \\
\hline
combine & 9.54 & 5.23\\
\hline
\end{tabular}
\end{table}

\subsection{Single or multiple \p\ formation scenario}\label{sec:discussion:multiple}

In this study, we have assumed that a single \p\ stars is born within
a halo. However, recent radiative hydrodynamical simulations showed
that massive \p\ binary stars can be born as a result of fragmentation of
circumstellar disks in the vicinity of \p\ due to the gravitational
instability \citep{Sugimura2020}. The total mass of individual stars
in their simulations is consistent with single star simulations
\citep{Sugimura2023}.  Therefore, the \p\ mass predicted by our model
can be interpreted as an upper limit of that  per halo.

Another possibility is the formation of \p\ clusters as shown by
radiative hydrodynamical simulations \citep{Hirano2018,Hirano2023}. As
the value of initial streaming velocity increases, the morphology of
the massive gas cloud in a halo becomes more asymmetric; hence, the
fragmentation of massive gas can occur more easily.  In their
simulations, the maximum of four dense gas clouds can form
simultaneously.

A cosmological magnetic field can also impact the formation of
\p\ stars.  Magnetohydrodynamic simulations showed that the dynamo
effect can exponentially amplify the cosmological magnetic field
strength around the \p\ protostars and then prevent disk
fragmentation, resulting in the formation of single massive star
\citep{Hirano2022,Higashi2024}.

Overall, there is no consensus about the multiplicity of the
\p\ formation at this moment, and therefore, we have assumed that a single
\p\ star is born within a halo throughout this work.  We will
investigate this issue in our future work.

\subsection{Caveats associated with the second peak of the \p\ IMF and SFRD}\label{sec:discussion:caveats}

As mentioned in \S~\ref{sec:imf} and~\ref{sec:sfrd}, 
there are two peaks of \p\ mass function and SFRD, corresponding to 
the H$_2$ ($20 \lesssim z \lesssim 25$) and atomic cooling halos ($z \lesssim 15$). 
These double peaks have not been reported in any literature partially 
due to insufficient computational volume or spatially uniform LW background.
We further discuss other possible caveats associated the second peak: 
external metal enrichment, photo-ionizing feedback, and X-ray feedback, 
which are not accounted in our current model.

Supernova explosions of stars born in halos enrich the gas inside the
halos with metals, which is the internal metal enrichment
process. Outflows can go out beyond the virial radius of the halos and
would enrich the surrounding gas and other nearby halos, which is the
external metal enrichment process.  Our model includes only the
internal enrichment by \p\ stars, while some semi-analytic models
include also the external enrichment by \p\ and \pii\ stars
\citep[e.g.,][]{Visbal2018,Visbal2020,Hartwig2022}.  The external
enrichment is negative feedback for both H$_2$ and atomic cooling
halos and should reduce the overall \p\ SFRD.

As well as the Lyman-Werner photodissociation, photo-ionizing radiation produced by
stars photoheats the surrounding gas and would suppress
star formation in relatively small halos.  The characteristic halo
mass at which the half of the baryon escapes by this process is estimated to be around
$10^7$\hMsun\ at $z=10$ \citep[e.g.,][]{Okamoto2008,Mutch2016}, which
is slightly less massive than the typical halo mass where supermassive
stars form in our model at $z<10$, $10^7 \lesssim M_{\rm vir} \lesssim
10^8$\hMsun, as seen in Figure~\ref{fig:cor}.

X-rays and cosmic-rays from X-ray binaries in nearby star-forming galaxies 
could enhance the H$_2$ formation in halos  via electron-catalyzed reaction 
and decrease the number of direct collapse halos \citep[e.g.,][]{Inayoshi2011}. 
\citet{Kimura2025} examined the effect of X-rays on direct collapse halos by a semi-analytic model 
and showed that the direct collapse is strongly suppressed by strong X-ray, leading H$_2$ cooling. 

Our model implements the Lyman-Werner feedback in a self-consistent way, 
however, does not consider these three feedback. 
These feedback can also be implemented in the self-consistent way, 
and we plan to investigate these effects in future work.

%%%%%%%%%%%%%%%%%%%%%%%%%%%%%%%%%%%%%%%%%%%%%%%%%%%%%%%%%%%%%%%%%%%%%%%%%%%%%%
%%%%%%%%%%%%%%%%%%%%%%%%%%%%%%%%%%%%%%%%%%%%%%%%%%%%%%%%%%%%%%%%%%%%%%%%%%%%%%
%%%%%%%%%%%%%%%%%%%%%%%%%%%%%%%%%%%%%%%%%%%%%%%%%%%%%%%%%%%%%%%%%%%%%%%%%%%%%%
%%%%%%%%%%%%%%%%%%%%%%%%%%%%%%%%%%%%%%%%%%%%%%%%%%%%%%%%%%%%%%%%%%%%%%%%%%%%%%
\section{Summary}\label{sec:summary}

In this paper, we have developed a new semi-analytic framework of
\p\ stars and subsequent galaxy formation designed to run on halo
merger trees.  In our framework, the critical halo mass for the
\p\ formation depends on the LW flux from \pandii\ sources and
the dark matter baryon streaming velocity, motivated by results of
recent radiative hydrodynamical simulations.  The \p\ mass is also
modelled as reproducing simulation results, taking the formation of
supermassive stars into account.  Our model incorporates the LW
feedback in a self-consistent way, and therefore, the spatial variation of
LW feedback naturally emerges.  We also conducted a high-resolution 
cosmological simulation with the box size of 16\hMpc\ having enough mass
resolution to resolve \p-forming halos, which is a much larger
volume than that adopted in previous studies using semi-analytic
models and radiative hydrodynamical simulations.  We have combined
this simulation with the semi-analytic framework and investigated the
\p\ IMF and SFRD of \pandii\
from $z=43$ to 7.5.  The results are summarized as follows.

\begin{enumerate}
\item 
The IMF of \p\ stars is top-heavy and two peaks in
the distribution exist.  The first peak is around the less massive end,
which corresponds to the H$_2$ cooling halos, while the second peak
corresponds to the atomic cooling halos, where supermassive stars
can be born.  The \p\ mass at the first peak increases with the value of
streaming velocity because the critical halo mass of \p-forming halos
also increases. The mass of the second peak is $\sim 2 \times 10^5$\Msun\ in
our model regardless of the value of streaming velocity.  The
fractions of such supermassive stars are about 2, 21, 57, and 83\% for
$v_{\rm bc}=0$, 1\sv, 2\sv, and 3\sv, respectively.
\item 
Regardless of the value of streaming velocity, the corresponding
redshift and LW flux of the first peak on the IMF are around 
$20 \lesssim z \lesssim 25$ and $J_{\rm LW} \lesssim 10$, 
respectively.  The second peak exists around 
$7.5 \lesssim z \lesssim 15$ and $J_{\rm LW} \gsim 10$.  
As the value of streaming velocity increases, the second peaks
emerge from slightly higher redshift and the number density of \p\ therein
also increases.
\item 
The \p\ formation begins around $z\sim40$, and the formation rate (in
terms of the number of \p-forming halos) rises with evolving
redshift and reaches the first peak at $20 \lesssim z \lesssim 25$.
Around this redshift, LW flux sufficiently grows to prevent subsequent
\p\ formation.  The formation rate rises again and reaches the second peak
around $7.5 \lesssim z \lesssim 15$ as new \p\ stars are born in
atomic cooling halos, which is also seen in SFRD.  
The \p\ SFRD is higher with
increasing streaming velocity, while the formation rate density of
\p-forming halos shows the opposite trend in the H$_2$ cooling regime.
In contrast, in the atomic cooling regime, both the SFRD and
\p-forming halo formation rate are higher with increasing streaming
velocity.
\item 
To incorporate LW feedback sufficiently from surrounding \pii\ stars 
and capture the atomic cooling regime,
at least an 8\hMpc\ simulation box is necessary in the case with the streaming
velocity, while even a larger box is necessary in the case without the
streaming velocity.  A model adopting spatially uniform LW background
used in many previous literature can not capture the atomic cooling
regime, therefore, the self-consistent model used in this paper is
necessary to correctly model \p\ stars in the atomic cooling halos.
\item
Our model predicts one supermassive star per halo with several
$10^9$\Msun\ at z=$7.5$, which is enough to reproduce a high redshift
quasar.
\end{enumerate}

%%%%%%%%%%%%%%%%%%%%%%%%%%%%%%%%%%%%%%%%%%%%%%%%%%%%%%%%%%%%%%%%%%%%%%%%%%%%%%
%%%%%%%%%%%%%%%%%%%%%%%%%%%%%%%%%%%%%%%%%%%%%%%%%%%%%%%%%%%%%%%%%%%%%%%%%%%%%%
%%%%%%%%%%%%%%%%%%%%%%%%%%%%%%%%%%%%%%%%%%%%%%%%%%%%%%%%%%%%%%%%%%%%%%%%%%%%%%
%%%%%%%%%%%%%%%%%%%%%%%%%%%%%%%%%%%%%%%%%%%%%%%%%%%%%%%%%%%%%%%%%%%%%%%%%%%%%%
\begin{acknowledgments}
This work has been supported by IAAR Research Support Program in Chiba University Japan,
MEXT/JSPS KAKENHI (Grant Number JP17H04828, JP19KK0344, JP21H01122 (T.I), JP21K13960, JP21H01123 (S.H)), 
MEXT as ``Program for Promoting Researches on the Supercomputer Fugaku''
(JPMXP1020200109 and JPMXP1020230406), and JICFuS.
Numerical computations were carried out on Aterui-II supercomputer at the Center
for Computational Astrophysics, National Astronomical Observatory of
Japan.
\end{acknowledgments}

\bibliographystyle{aasjournal}

%%%%%%%%%%%%%%%%%%%%%%%%%%%%%%%%%%%%%%%%%%%%%%%%%%%%%%%%%%%%%%%%%%%%%%%%%%%%%%
%%%%%%%%%%%%%%%%%%%%%%%%%%%%%%%%%%%%%%%%%%%%%%%%%%%%%%%%%%%%%%%%%%%%%%%%%%%%%%
%%%%%%%%%%%%%%%%%%%%%%%%%%%%%%%%%%%%%%%%%%%%%%%%%%%%%%%%%%%%%%%%%%%%%%%%%%%%%%
%%%%%%%%%%%%%%%%%%%%%%%%%%%%%%%%%%%%%%%%%%%%%%%%%%%%%%%%%%%%%%%%%%%%%%%%%%%%%%
\appendix
\restartappendixnumbering

\section{Comparison of \mcrit}\label{sec:mcrit_comp}

\begin{figure*}
\centering
\includegraphics[width=0.45\linewidth]{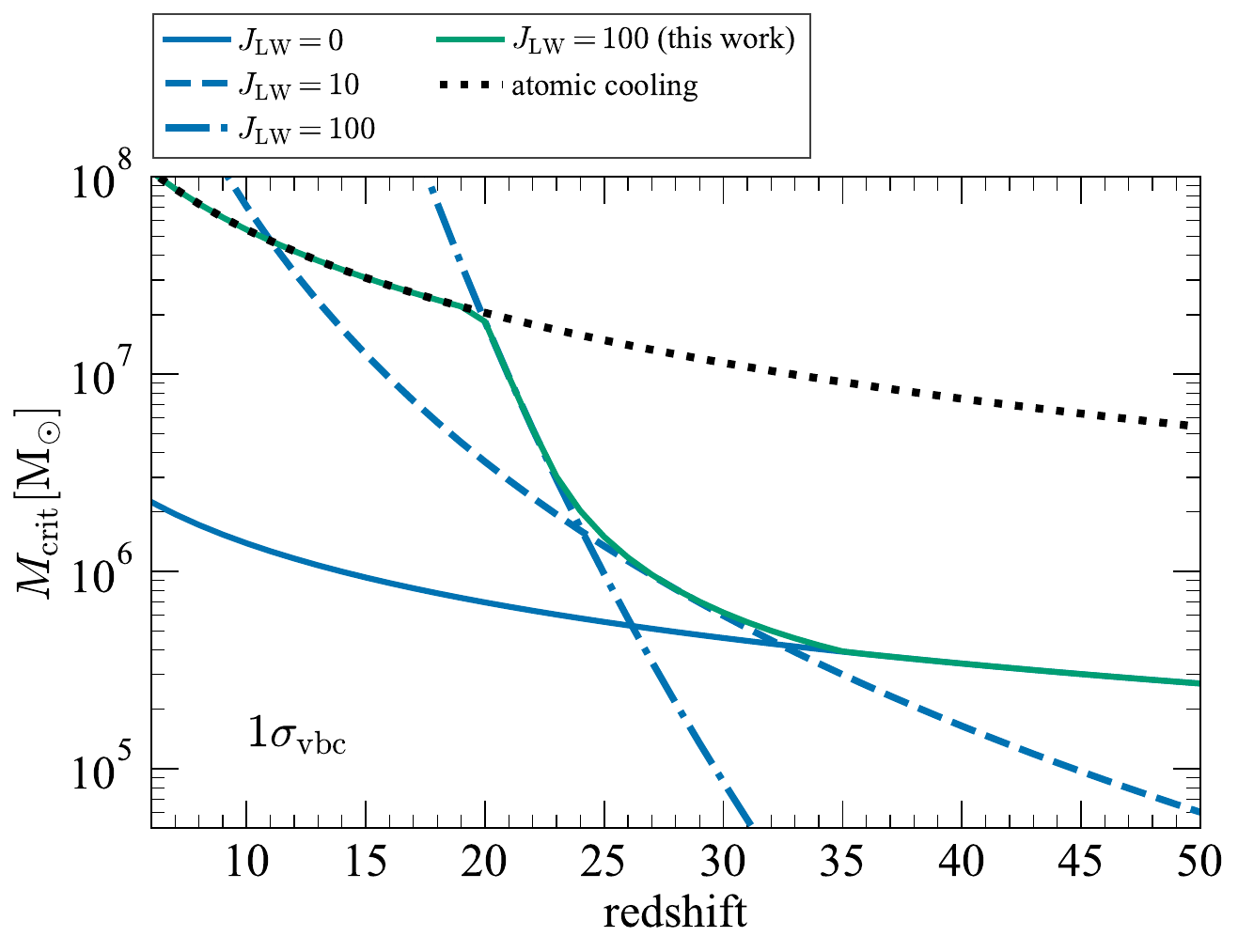}
\includegraphics[width=0.45\linewidth]{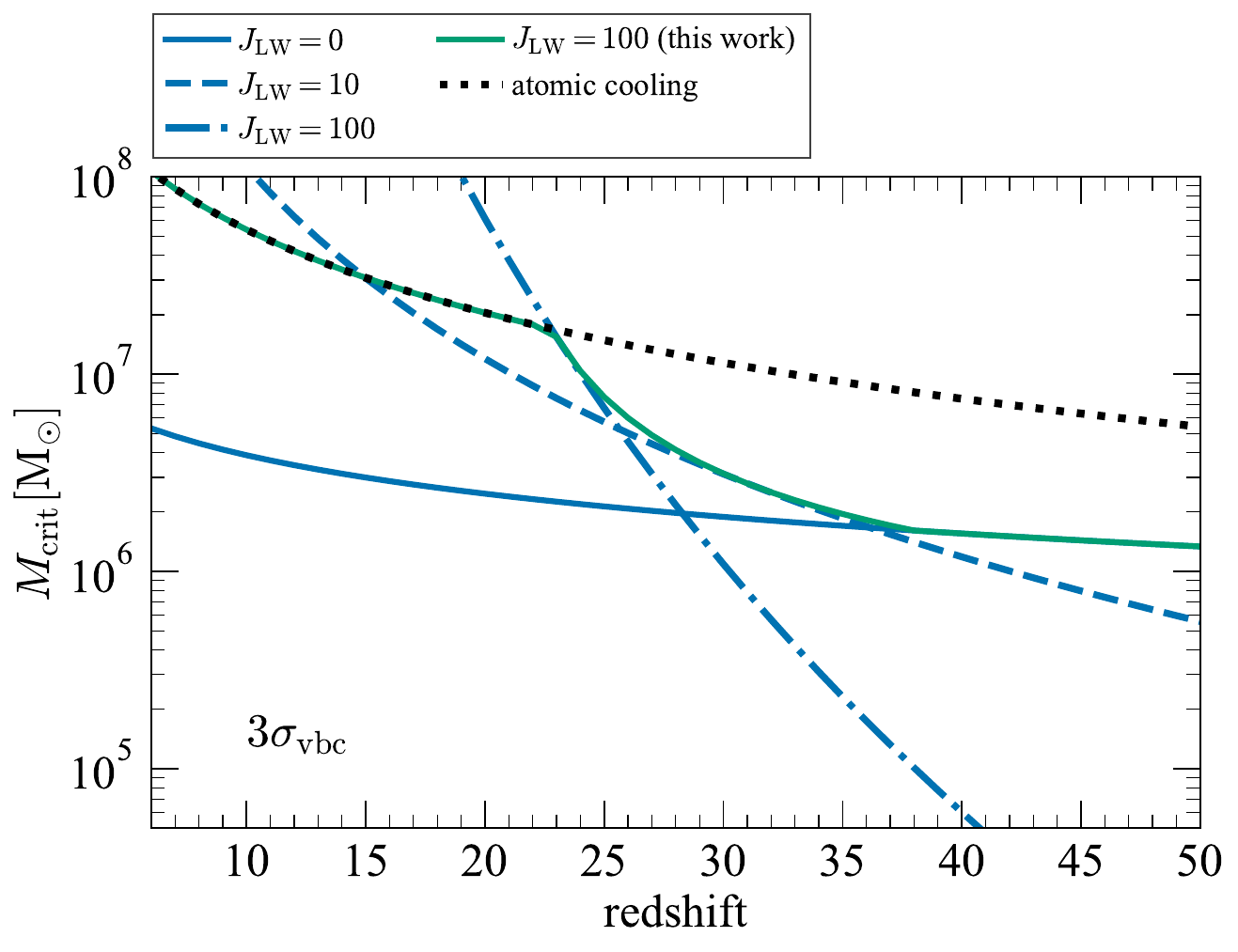}
\caption{
Comparison of critical halo mass \mcrit\ of \p-forming halos as a function of redshift employed in our semi-analytic model [Eq~\eqref{eq:mcrit}] 
(solid green curve) 
and one originally proposed by \citet{Kulkarni2021}
(three blue curves). 
Blue solid, dashed, and dot-dashed curves indicate 
the results of $J_{\rm LW}=0, 10$, and 100, respectively.
Upward thick curves represent the critical mass of atomic cooling halos [Eq~\eqref{eq:atomic_cooling}].
Left and right panels show the results with the streaming velocity of 1\sv\ and 3\sv, respectively.
}
\label{fig:mcrit_comp}
\end{figure*}

As described in \S~\ref{sec:model:pop3}, 
the critical halo mass \mcrit\ of \p-forming halos employed in our semi-analytic model [Eq~\eqref{eq:mcrit}] is based on a fitting function proposed by \citet{Kulkarni2021}, 
which depends on the value of redshift, LW flux, and streaming velocity.
However, this fitting function gives nonphysical results when it is adopted to outside 
of parameter spaces used in the calibration. 
Figure~\ref{fig:mcrit_comp} shows the original fitting function 
proposed by \citet{Kulkarni2021} for the streaming velocity of 1\sv\ and 3\sv. 
In both cases, the critical mass decreases with increasing LW flux at high redshift ($z \gtrsim 25$). This is physically strange but is not surprising because
this fitting function was not calibrated for such high \jlw. To apply
this fitting function to such parameter spaces, we replace it with the prescription shown in 
[Eq~\eqref{eq:kulkarni21b}] as the critical halo mass increases monotonically with \jlw.

%%%%%%%%%%%%%%%%%%%%%%%%%%%%%%%%%%%%%%%%%%%%%%%%%%%%%%%%%%%%%%%%%%%%%%%%%%%%%%
%%%%%%%%%%%%%%%%%%%%%%%%%%%%%%%%%%%%%%%%%%%%%%%%%%%%%%%%%%%%%%%%%%%%%%%%%%%%%%
%%%%%%%%%%%%%%%%%%%%%%%%%%%%%%%%%%%%%%%%%%%%%%%%%%%%%%%%%%%%%%%%%%%%%%%%%%%%%%
%%%%%%%%%%%%%%%%%%%%%%%%%%%%%%%%%%%%%%%%%%%%%%%%%%%%%%%%%%%%%%%%%%%%%%%%%%%%%%

\section{Effect of mass resolution}\label{sec:mass_convergence}

\begin{figure*}[b]
\centering
\includegraphics[width=0.368\linewidth]{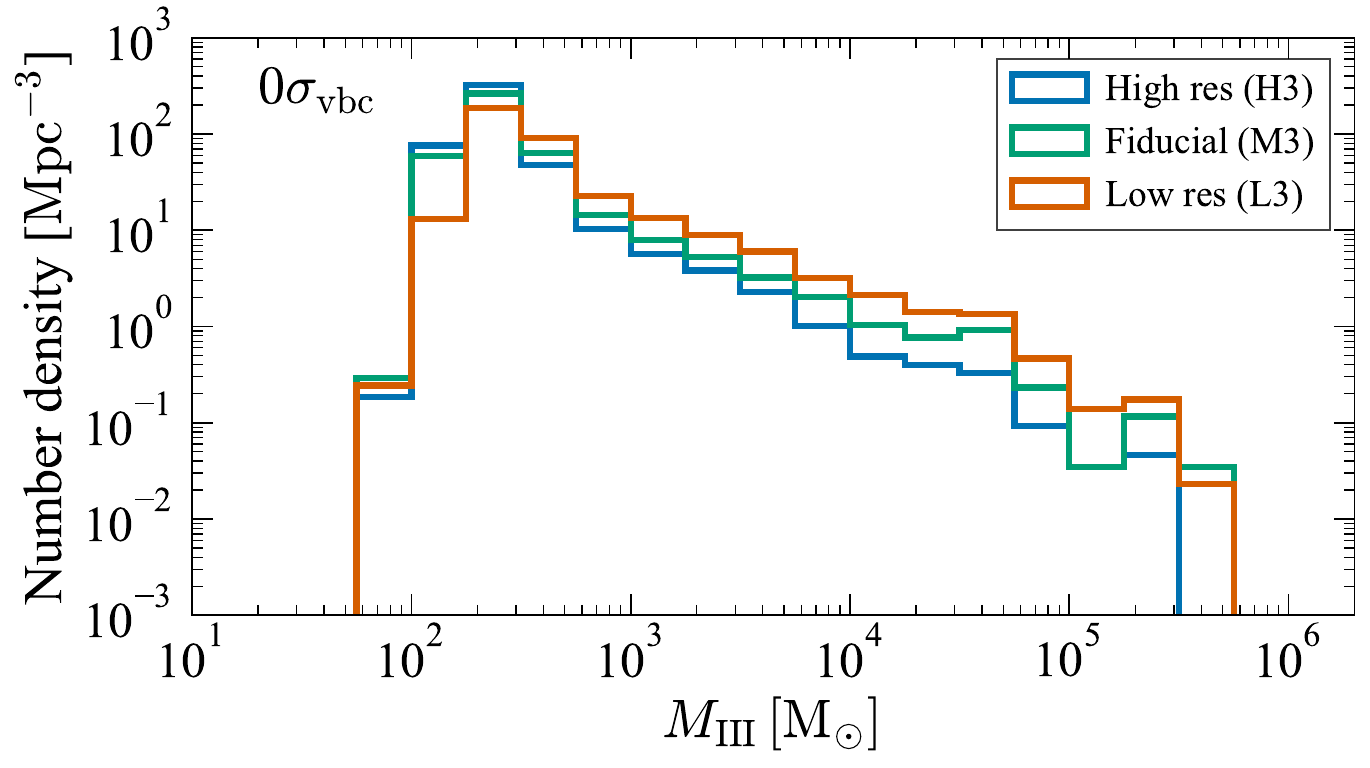}
\includegraphics[width=0.368\linewidth]{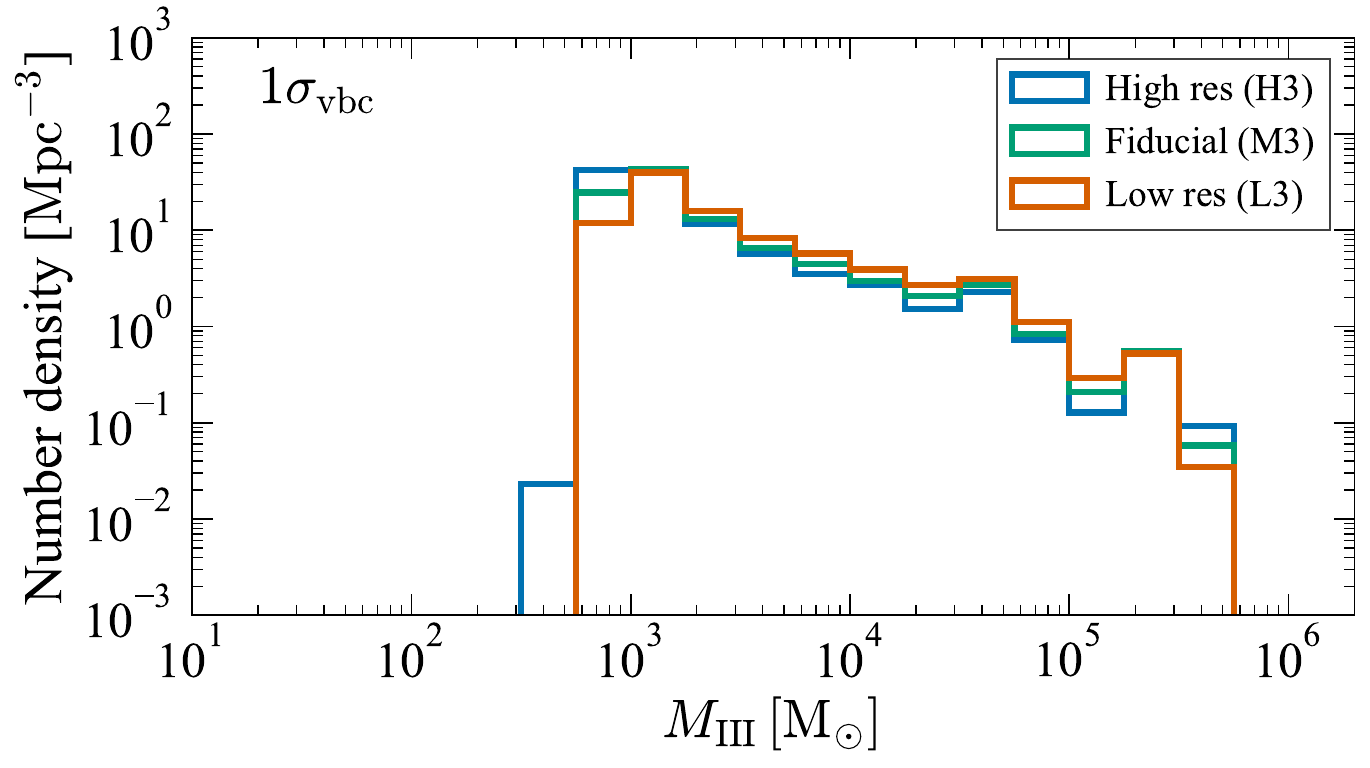}
\includegraphics[width=0.368\linewidth]{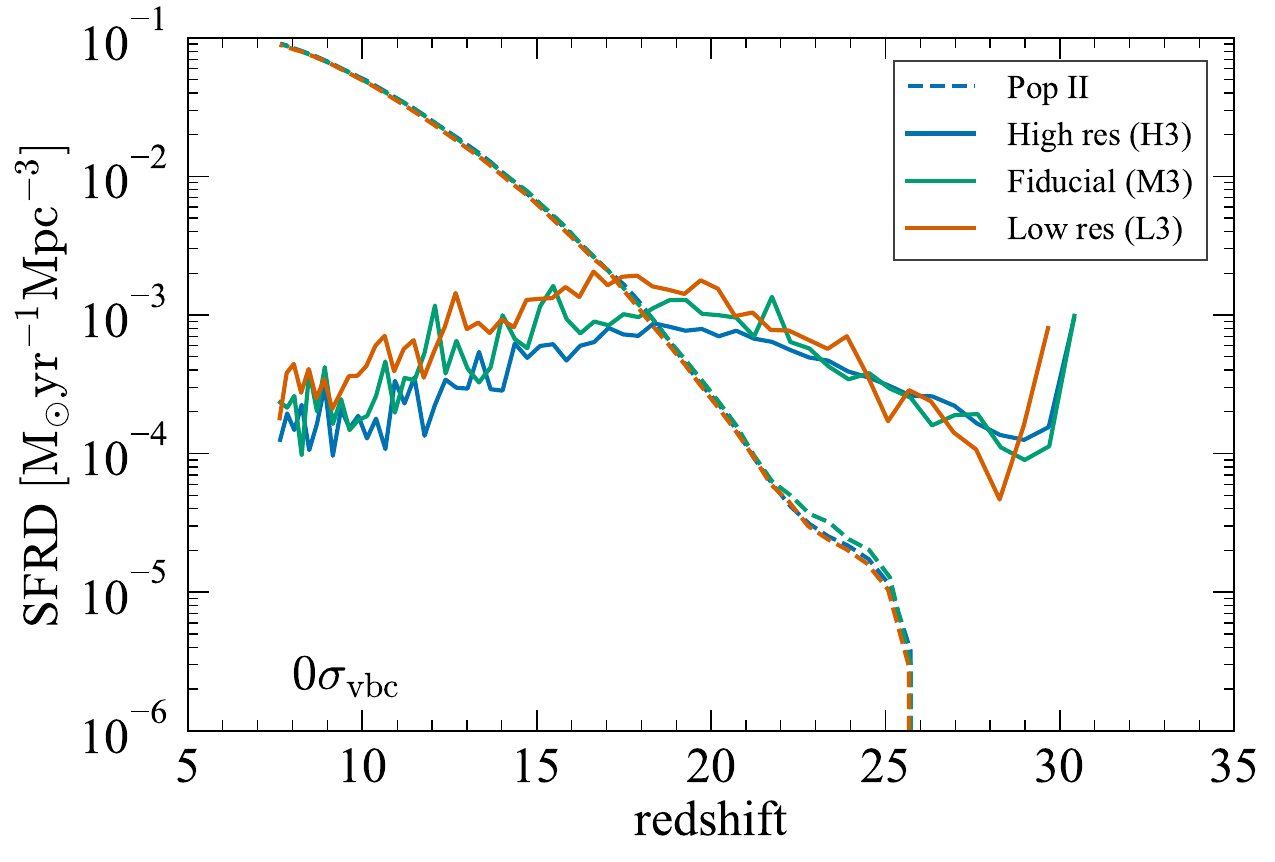}
\includegraphics[width=0.368\linewidth]{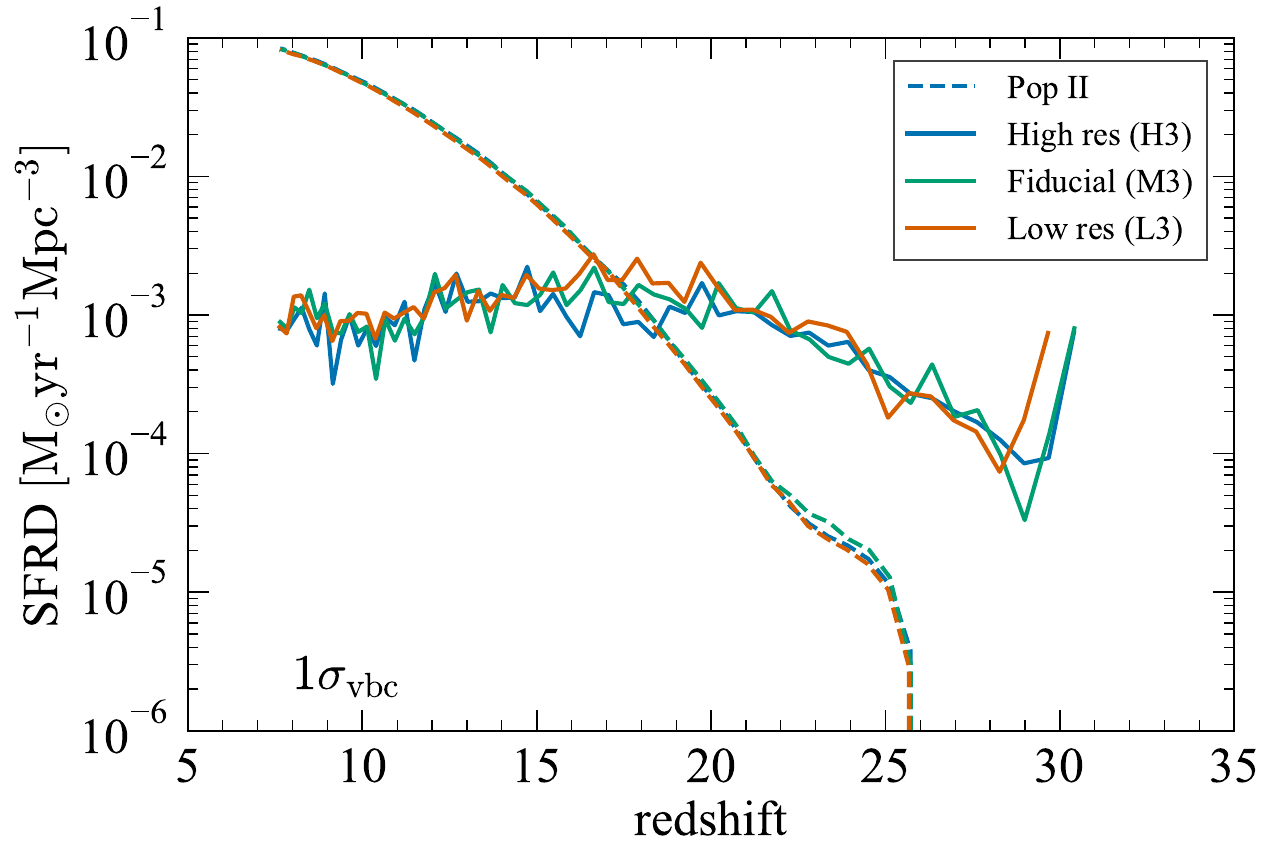}
\includegraphics[width=0.368\linewidth]{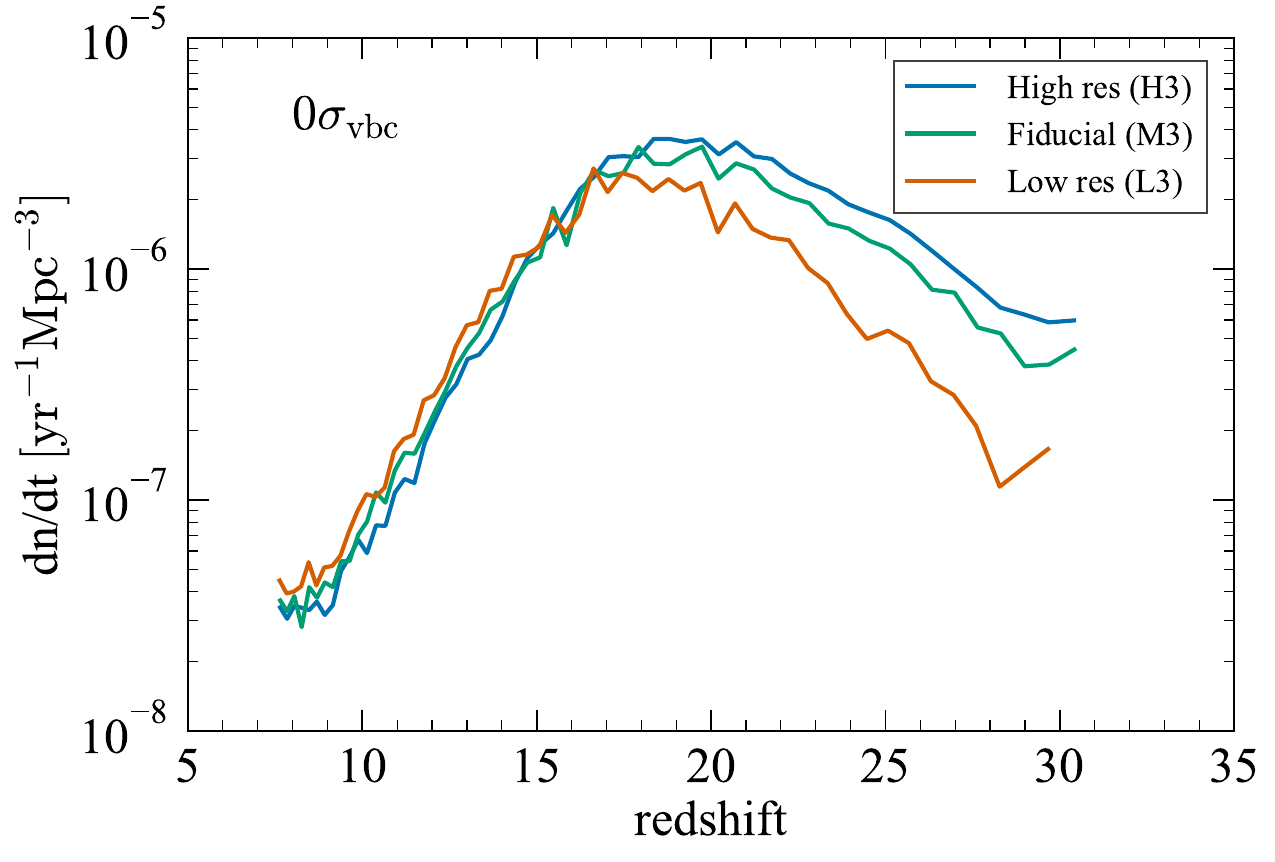}
\includegraphics[width=0.368\linewidth]{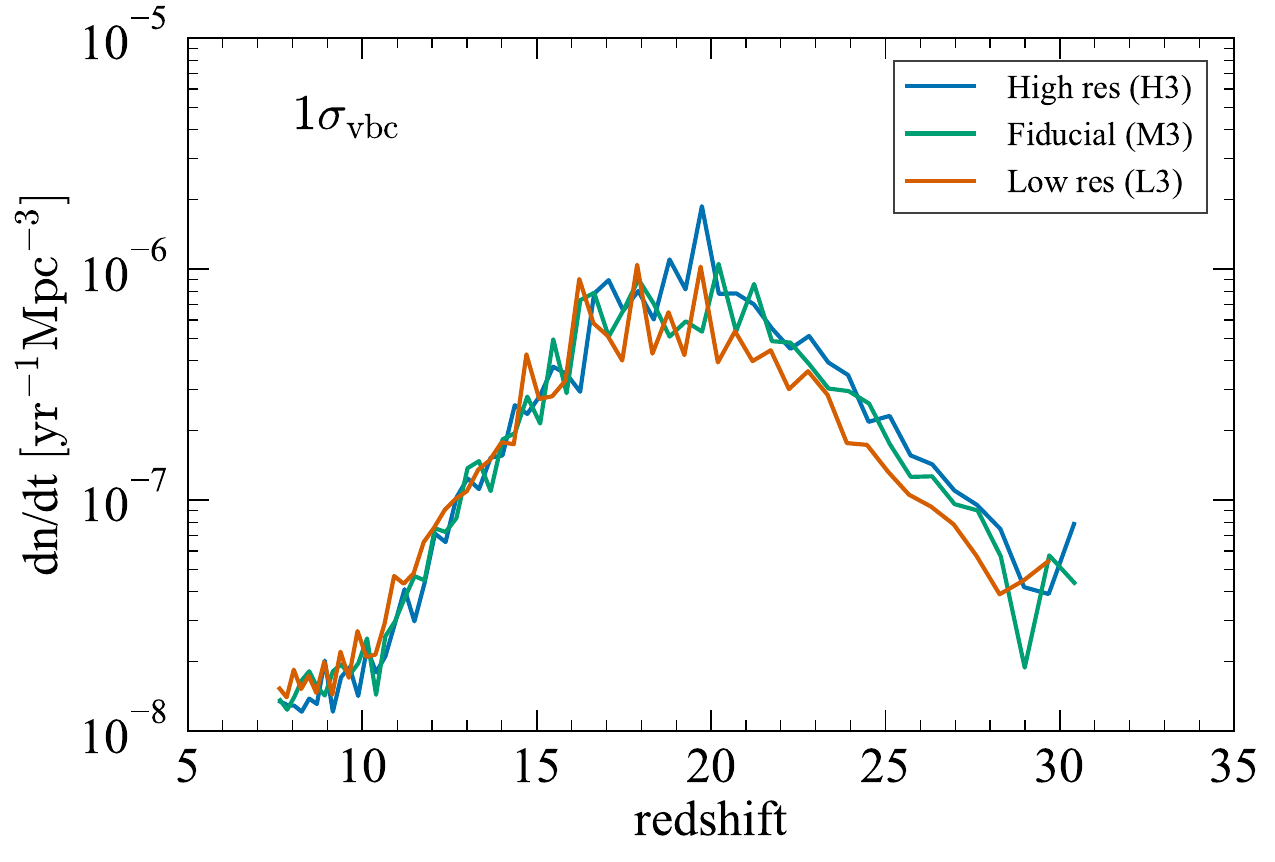}
\caption{
Comparison of model results calculated on simulations 
with different mass resolutions
(\Ht, \Mt, and \Lt).
Left and right three panels show results of model with 
$v_{\rm bc} = 0$ and 1\sv, respectively. 
(Top) Number density of \p\ stars formed at all redshift as a function of the
\p\ mass in a halo, same as Figure~\ref{fig:imf}.  (Middle) Star
formation rate density, same as Figure~\ref{fig:sfrd}.  (Bottom)
Formation rate density of \p-forming halos, same as
Figure~\ref{fig:sfrd}.  
}
\label{fig:mres_conv}
\end{figure*}

In this appendix, we examine the effect of the mass resolution of
simulations on the IMF of \p\ stars formed at all redshift and
SFRD.  Figure~\ref{fig:mres_conv} shows the comparison of model
results calculated on simulations with different mass resolutions (\Ht,
\Mt, and \Lt), which are $6.41 \times 10^{2}$, $5.13 \times 10^{3}$,
and $1.73 \times 10^{4}$\hMsun.  The resolution of \Mt\ is the same
as the fiducial resolution, while that of \Lt\ is the lowest and
similar to those used in other semi-analytic studies \citep{Magg2018,
  Griffen2018}.  The results with $v_{\rm bc} = 0$ (left panels) and
1\sv\ (right panels) are shown. 

For the 1\sv\ case, three simulations give nearly
converged results. For the $v_{\rm bc} = 0$ case, the \Ht\ and
\Mt\ simulations agree with each other, although the middle-resolution
simulation slightly overpredicts \p\ stars at the massive end.  On the
other hand, the \Lt\ run underpredicts the formation rate of halos 
at $z \gtrsim 15$ and slightly  overpredicts it at $z \lesssim 15$,
indicating that the mass resolution of \Lt\ is not enough to resolve
halos  with the mass close to the smallest halo mass 
($\sim 10^5$\hMsun, see Figure~\ref{fig:mcrit}). Because the mass of such
halos  exceeds the critical mass at a late time as they grow, the
formation rate of halos  at $z \lesssim 15$ is slightly larger in lower
resolution than in higher resolution simulations.  As a result, the
\Lt\ run gives a larger number of massive stars and higher SFRD
compared to the other runs for the $v_{\rm bc} = 0$ case.  For the $v_{\rm
  bc} = 1\sigma_{\rm vbc}$ case, this does not occur because the
critical halo mass is always higher.

%%%%%%%%%%%%%%%%%%%%%%%%%%%%%%%%%%%%%%%%%%%%%%%%%%%%%%%%%%%%%%%%%%%%%%%%%%%%%%
%%%%%%%%%%%%%%%%%%%%%%%%%%%%%%%%%%%%%%%%%%%%%%%%%%%%%%%%%%%%%%%%%%%%%%%%%%%%%%
%%%%%%%%%%%%%%%%%%%%%%%%%%%%%%%%%%%%%%%%%%%%%%%%%%%%%%%%%%%%%%%%%%%%%%%%%%%%%%
%%%%%%%%%%%%%%%%%%%%%%%%%%%%%%%%%%%%%%%%%%%%%%%%%%%%%%%%%%%%%%%%%%%%%%%%%%%%%%
\end{document}